\documentclass[aps,prd,reprint,superscriptaddress,floatfix,longbibliography,showpacs=false]{revtex4-2}

\usepackage[utf8]{inputenc}
\usepackage[T1]{fontenc}
\usepackage{amsmath,amssymb,amsfonts}
\usepackage{graphicx}
\usepackage{xcolor}
\usepackage{booktabs}
\usepackage[colorlinks=true,linkcolor=blue!50!black,citecolor=blue!50!black,urlcolor=blue!50!black]{hyperref}

\newcommand{\msun}{\ensuremath{\mathrm{M_\odot}}}
\newcommand{\kpc}{\ensuremath{\mathrm{kpc}}}
\newcommand{\dsph}{dSph}

\begin{document}

\title{Hydrostatic Equilibrium Estimate of the Fuzzy Dark Matter Mass from Dwarf Spheroidals: Method and Systematics}

\author{Samrat Mitra}
\email{sm26rs015@iiserkol.ac.in}
\affiliation{Department of Physical Sciences, Indian Institute of Science Education and Research (IISER) Kolkata, Mohanpur 741246, West Bengal, India}
\author{Koushik Dutta}
\email{koushik@iiserkol.ac.in}
\affiliation{Department of Physical Sciences, Indian Institute of Science Education and Research (IISER) Kolkata, Mohanpur 741246, West Bengal, India}

\begin{abstract}
We present an estimator of an effective fuzzy (wave) dark matter (FDM) boson mass based on local hydrostatic equilibrium, which requires no soliton or other parametric density template. The condition requires the classical gravitational acceleration to be balanced by the gradient of the quantum potential in the Schr\"odinger--Poisson--Madelung system. We apply it to the density and enclosed-mass posteriors of nine Milky Way dwarf spheroidal (dSph) galaxies obtained with \textsc{pyGravSphere}. For each posterior realization, we determine by least squares the boson mass that best satisfies the force-balance condition over a fixed radial window. This gives a distribution of \(m_{21}\equiv m/10^{-21}\,\mathrm{eV}\) for each galaxy. The per-galaxy median masses range from \(m_{21}\sim 8\) to \(46\), corresponding to \(m\sim(0.8\text{--}4.6)\times10^{-20}\,\mathrm{eV}\); Draco and Ursa Minor have the lowest median masses, while Carina and Sextans have the highest. The inverse-variance-weighted effective equilibrium mass of the nine galaxies is \(m=(1.31\pm0.35)\times10^{-20}\,\mathrm{eV}\), with Draco and Ursa Minor contributing most of the weight.  We also examine the systematic uncertainties in our results.
\end{abstract}

\maketitle

\section{Introduction}

The nature of dark matter (DM) remains one of the central open problems in cosmology and astroparticle physics. Although the cold dark matter (CDM) paradigm has been highly successful in describing structure formation on large scales, it faces several long-standing challenges on galactic and sub-galactic scales, including the core-cusp problem, the missing satellites problem, and the too-big-to-fail problem  \cite{BullockBoylanKolchin2017}. Fuzzy (or wave) dark matter (FDM), in which DM is composed of an ultralight bosonic field with mass $m \sim 10^{-22}$--$10^{-21}\,\mathrm{eV}$, has been proposed as an alternative that can suppress small-scale structure and naturally produce kiloparsec-scale cores through macroscopic quantum pressure  \cite{Hu2000,Hui2017,Hui2021}. On scales below the de Broglie wavelength, the dynamics of such a field are governed by the coupled Schrödinger-Poisson (SP) equations, introduced as an effective description of ultralight bosonic DM by Sin  \cite{Sin1994} and Widrow and Kaiser  \cite{Widrow1993}. These equations admit stable, self-gravitating ground-state solutions known as solitons, which can form dynamically through a dissipationless process called gravitational cooling  \cite{SeidelSuen1994,GuzmanUrenaLopez2006}. Cosmological and idealized simulations have shown that such solitons can persist at the centres of FDM halos  \cite{Schive2014a,Schive2014b,Schwabe2016,Mocz2017,Veltmaat2018,MaySpringel2021,Nori2019,Chakrabarti:2022owq}. The soliton--halo connection has subsequently been used
to constrain the boson mass $m$ using galactic rotation
curves  \cite{Bar2018,BarBlumEbySato2019} and mixed
dark matter halo formation models
 \cite{MarshSilk2014,AmendolaBarbieri2006}. Related studies have also investigated the relaxation and stability of relativistic axion-star configurations using direct Klein-Gordon simulations  \cite{Levkov2017}.

The characteristic role of quantum pressure in supporting an FDM core depends on the boson mass: the coefficient of the quantum-pressure term scales as $m^{-2}$. The central density structure and gravitational potential of a DM-dominated galaxy can therefore provide an astrophysical probe of the DM mass. Dwarf spheroidal (dSph) satellites of the Milky Way are particularly well suited to this purpose. They are strongly DM dominated, are generally less affected by baryonic feedback than more massive galaxies, and have extensive, resolved stellar-kinematic datasets. These observations allow their DM density and enclosed-mass profiles to be reconstructed using Jeans-equation-based dynamical modelling  \cite{Walker2009,Mateo1998}. The reconstructed profiles provide an opportunity to investigate the role of quantum pressure in the internal gravitational structure of these galaxies and to test whether their inferred density and mass distributions are compatible with the equilibrium condition of the SP system.

A substantial body of work has used dSph kinematics to constrain the FDM boson mass, although the resulting bounds and preferred mass ranges differ, sometimes by factors of several. Early estimates based on core-radius and central-density arguments, together with mock-data studies, highlighted the possibility of biases arising from the degeneracy between the DM mass distribution and stellar velocity anisotropy  \cite{GonzalezMorales2017}. Subsequent studies have adopted a variety of approaches, including full wave-DM Jeans analyses  \cite{Chen2017}, soliton-halo scaling relations  \cite{Bar2018}, halo-mass-dependent soliton-density models calibrated against cosmological simulations  \cite{RoblesBullockBoylanKolchin2019}, non-spherical halo modelling  \cite{HayashiObata2020}, comparisons of density-profile diversity across the dSph population  \cite{HayashiChibaIshiyama2020}, dedicated analyses of ultra-faint dwarfs  \cite{MarshNiemeyer2019,HayashiFerreiraChan2021}, and dynamical-heating arguments  \cite{Safarzadeh2020,Church2019}. Complementary constraints arise from probes on larger cosmological scales, including the Lyman-$\alpha$ forest power spectrum  \cite{Irsic2017,Kobayashi2017,Armengaud2017,RogersPeiris2021}, the abundance and luminosity function of Milky Way satellites  \cite{Nadler2021,Schutz2020}, high-redshift galaxy UV luminosity functions during reionization  \cite{Bozek2015}, and precision CMB and large-scale-structure observations  \cite{Hlozek2015}. These studies collectively disfavor the lightest end of the FDM mass range originally proposed to address small-scale structure problems; see Refs.  \cite{Ferreira2021,Hui2021} for comprehensive reviews. More recently, Dalal and Kravtsov  \cite{Dalal2022} used the sizes and kinematics of ultra-faint dwarfs to exclude $m \lesssim 3\times10^{-19}\,\mathrm{eV}$. Zimmermann \textit{et al.}  \cite{Zimmermann2025} combined a model-independent, non-parametric reconstruction of the Leo II phase-space distribution with a maximum mean discrepancy statistic to derive a lower bound of $m > 2.2\times10^{-21}\,\mathrm{eV}$ at $95\%$ confidence.

The reliability of dSph-based constraints depends critically on the reconstruction of the underlying DM density and mass profiles from stellar kinematic data. In particular, the mass-velocity-anisotropy degeneracy can introduce significant uncertainties if the dynamical modelling relies on restrictive assumptions about the density profile or stellar velocity anisotropy. This motivates the use of non-parametric, higher-order Jeans modelling frameworks, such as \textsc{GravSphere}  \cite{ReadSteger2017,ReadDraco2018,Genina2020}, which incorporate higher-order kinematic information to better account for this degeneracy. Such methods provide posterior distributions for the density and enclosed-mass profiles rather than a single best-fitting parametric model. Propagating these posterior distributions is important for assessing the uncertainties in any subsequent inference of the FDM boson mass.

In this work, we adopt a different strategy to constrain the boson mass. Rather than fitting a prescribed soliton or soliton-plus-NFW density profile to the reconstructed DM distribution, or comparing the full phase-space distribution, we directly test the local hydrostatic-equilibrium condition of the SP system on the reconstructed density and enclosed-mass profiles. We ask what boson mass is \emph{required} for static SP hydrostatic equilibrium. In the Madelung formulation, this condition requires the classical gravitational acceleration to be balanced by the gradient of the quantum (Bohm) potential. For each posterior realization of the density and mass profiles, we calculate the corresponding quantum-potential contribution and gravitational acceleration, and determine the single boson mass that best satisfies this force-balance condition over the chosen radial interval. We obtain this mass through an ordinary least-squares minimization of the fractional force-balance residual across the radial range. Repeating this procedure over the posterior realizations yields a distribution of equilibrium masses for each galaxy, thereby propagating the uncertainties in the dynamical reconstruction into the mass inference.

A central feature of this approach is that it does not assume a particular functional form for the DM density profile and does not require the galaxy to host a fully relaxed soliton. The equilibrium condition can be evaluated directly for any reconstructed spherical density and enclosed-mass profile, without imposing the analytic soliton solution. Furthermore, unlike methods that infer the boson mass through a relation between soliton properties and the host halo, our estimator does not invoke a soliton-halo relation or require its calibration against cosmological simulations. It does not rely on a cosmological prescription for halo formation, a simulated halo-formation history, or a cosmological relation between the central soliton and the host halo. The inference therefore uses the present-day dynamical profiles of the observed galaxies and is independent of cosmological assumptions about how their halos formed and assembled. The estimator also does not require cosmological parameters such as $\Omega_{\rm DM}$, nor does it rely on constraints from the early Universe or large-scale structure. In this sense, our method provides a direct, template-free test of the local SP force-balance condition, subject to the assumed static and spherically symmetric regime, without relying on cosmological predictions for the internal structure of FDM halos.

We apply this estimator to nine Milky Way dSphs. This paper presents the complete derivation of the method, its application to all nine galaxies, the statistical combination of the individual-galaxy results, and the numerical implementation. We also examine the robustness of the inferred masses through systematic tests, including their dependence on the radial fitting window and the correlation between the force-balance residual and the inferred boson mass.

The remainder of this paper is organized as follows. Section  \ref{sec:theory} reviews the SP-Madelung formalism and derives the local equilibrium condition. Section  \ref{sec:data} describes the data, while Section  \ref{sec:pygs} summarizes the \textsc{pyGravSphere} mass-modelling methodology used to obtain the density and mass posteriors. Section  \ref{sec:method} details the regression estimator and its numerical implementation. Section  \ref{sec:results} presents the density, mass, and quantum-potential profiles, the local-equilibrium diagnostic, the individual-galaxy $m_{21}$ posteriors, and the inverse-variance-weighted effective equilibrium mass. Section  \ref{sec:robustness} examines the force-balance residual, the convergence fraction, and the sensitivity to the fitting window. Finally, Section  \ref{sec:discussion} compares our results with previous studies, and Section  \ref{sec:conclusion} presents our conclusions.

\section{Theoretical framework}
\label{sec:theory}

\subsection{Action and the Einstein--Klein--Gordon system}
\label{sec:action}

We take FDM to be a canonical, minimally coupled real scalar field $\phi(x^\mu)$ of mass $m$, described by the covariant action   \cite{Dave:2023egr}
\begin{equation}
\begin{split}
    S[\phi,g_{\mu\nu}] = \int d^4x \, \sqrt{-g}\Big[ &\frac{M_{\rm Pl}^2}{2}\,R
    - \frac{1}{2}\,g^{\mu\nu}\partial_\mu\phi\,\partial_\nu\phi \\
    &- \frac{1}{2}\left(\frac{mc}{\hbar}\right)^{\!2}\phi^2 \Big] ,
\end{split}
    \label{eq:action}
\end{equation}
where $R$ is the Ricci scalar of the spacetime metric $g_{\mu\nu}$, $M_{\rm Pl}^2 = (8\pi G)^{-1}$, and the second term is the canonical free field kinetic term while the third is the mass term of a real, non-self-interacting Klein-Gordon field  \cite{Hu2000,Hui2017,Marsh2016,Ferreira2021,Hui2021}. Varying Eq. \eqref{eq:action} with respect to $\phi$ gives the covariant Klein-Gordon equation,
\begin{equation}
    \frac{1}{\sqrt{-g}}\,\partial_\mu\!\left(\sqrt{-g}\,g^{\mu\nu}\partial_\nu\phi\right)
    - \left(\frac{mc}{\hbar}\right)^{\!2}\phi = 0 ,
    \label{eq:KG}
\end{equation}
while varying with respect to $g_{\mu\nu}$ gives the Einstein field equations sourced by the stress-energy tensor of $\phi$, $T_{\mu\nu} = \partial_\mu\phi\,\partial_\nu\phi - g_{\mu\nu}\left[\tfrac{1}{2}g^{\alpha\beta}\partial_\alpha\phi\partial_\beta\phi + \tfrac{1}{2}(mc/\hbar)^2\phi^2\right]$. Equations \eqref{eq:action}- \eqref{eq:KG} are the starting point of the entire FDM formalism used in this work; every equation below follows from them in a definite sequence of well controlled limits, which we now make explicit so that the density and mass normalization conventions used in our numerical pipeline are unambiguous.

\subsection{Weak-field, non-relativistic reduction to the Schr\"odinger--Poisson system}

We specialize to the regime relevant for a virialized dSph halo: weak gravity ($|\Phi|/c^2 \ll 1$) and non-relativistic bulk motion ($v/c \ll 1$), appropriate everywhere outside the immediate vicinity of a compact object. In the Newtonian-gauge, quasi-static metric
\begin{equation}
    ds^2 = -\left(1+\frac{2\Phi}{c^2}\right)c^2 dt^2 + \left(1-\frac{2\Phi}{c^2}\right)\delta_{ij}\,dx^i dx^j ,
\end{equation}
with $\Phi(\mathbf{r},t)$ the Newtonian potential, the Klein--Gordon equation \eqref{eq:KG} reduces, to leading order in $\Phi/c^2$, to
\begin{equation}
    \left(1-\frac{4\Phi}{c^2}\right)\frac{1}{c^2}\partial_t^2\phi
- \nabla^2\phi
+ \left(1-\frac{2\Phi}{c^2}\right)\left(\frac{mc}{\hbar}\right)^{\!2}\phi = 0 .,
    \label{eq:KGweak}
\end{equation}
while the Einstein equations reduce, in the same limit, to the Newtonian Poisson equation sourced by the (time-averaged) energy density of the field, $\nabla^2\Phi = 4\pi G\,\rho$.

In the nonrelativistic regime, the field $\phi$ contains a rapidly oscillating
rest-mass contribution at the Compton frequency
$\omega_c = \frac{mc^2}{\hbar}$,
whereas the spatial gradients, gravitational potential, and envelope time
variation provide only small corrections to this dominant rest-mass scale. We therefore separate the rapid Compton oscillation from the slowly varying nonrelativistic envelope by writing  \cite{Hu2000,Widrow1993,Hui2017}
\begin{equation}
\begin{split}
    \phi(\mathbf{r},t) = \frac{\hbar}{c\sqrt{2m}}\Big[&\psi_{\rm std}(\mathbf{r},t)\,e^{-i m c^2 t/\hbar} \\
    &+ \psi_{\rm std}^*(\mathbf{r},t)\,e^{+i m c^2 t/\hbar}\Big] ,
\end{split}
    \label{eq:WKB}
\end{equation}
with $\psi_{\rm std}$ a complex envelope assumed to vary slowly compared with the Compton time, $|\partial_t\psi_{\rm std}| \ll (mc^2/\hbar)|\psi_{\rm std}|$. Substituting Eq. \eqref{eq:WKB} into Eq. \eqref{eq:KGweak}, dropping $\partial_t^2\psi_{\rm std}$ (the non-relativistic approximation) and averaging over the fast Compton oscillation (which removes terms rotating at $e^{\pm 2imc^2t/\hbar}$) yields the Schr\"odinger equation
\begin{equation}
    i\hbar\,\partial_t \psi_{\rm std} = -\frac{\hbar^2}{2m}\nabla^2\psi_{\rm std} + m\Phi\,\psi_{\rm std} .
    \label{eq:Schrod-std}
\end{equation}
Carrying out the same time-average on the field's energy density, $\rho c^2 = \tfrac{1}{2}\left[(\partial_t\phi)^2 + c^2(\nabla\phi)^2 + (mc^2/\hbar)^2\phi^2\right]$, and substituting Eq. \eqref{eq:WKB}, the leading (non-oscillating) contribution is $\rho = m\,|\psi_{\rm std}|^2$  \cite{Hu2000,Hui2017}, so that the Poisson equation closes as
\begin{equation}
    \nabla^2\Phi = 4\pi G\, m\,|\psi_{\rm std}|^2 .
    \label{eq:Poisson-std}
\end{equation}
Equations  \eqref{eq:Schrod-std}-- \eqref{eq:Poisson-std} are the standard Schr\"odinger--Poisson (SP) system, in which $|\psi_{\rm std}|^2$ is a \emph{number} density and the mass density $\rho$ carries an explicit factor of $m$. This is an excellent approximation to the collisionless Vlasov--Poisson dynamics of classical CDM once wave interference effects are coarse-grained over scales larger than the de Broglie wavelength  \cite{Widrow1993,Hui2021}.

\subsection{Field normalization: \texorpdfstring{$\rho = |\Psi|^2$}{rho = |Psi|^2}}
\label{sec:normalization}

The SP system \eqref{eq:Schrod-std}- \eqref{eq:Poisson-std} is written above in the ``number-density'' convention common in the analytic literature, $\rho = m|\psi_{\rm std}|^2$. Our numerical pipeline instead adopts the convention that the wavefunction amplitude itself represents the mass density directly, $\rho \equiv |\Psi|^2$, with no explicit factor of $m$. The two conventions describe the same physics and differ only by an overall, time- and space-independent rescaling of the field amplitude; we make this explicit here so that every equation quoted in the remainder of the paper is manifestly consistent with the normalization used in the code.

Define the rescaled field $\Psi \equiv \sqrt{m}\,\psi_{\rm std}$, so that $|\Psi|^2 = m|\psi_{\rm std}|^2 = \rho$ is the mass density by construction. Since $m$ is a constant, dividing Eq. \eqref{eq:Schrod-std} by $\sqrt{m}$ leaves its functional form unchanged,
\begin{equation}
\begin{split}
    i\hbar\,\partial_t \Psi &= -\frac{\hbar^2}{2m}\nabla^2\Psi + m\Phi\,\Psi , \\
    \nabla^2\Phi &= 4\pi G\,\rho , \qquad \rho \equiv |\Psi|^2 ,
\end{split}
    \label{eq:SP}
\end{equation}
which is the Schr\"odinger--Poisson system we use throughout this work, with $\rho(\mathbf{r})=|\Psi(\mathbf{r})|^2$ read directly off the \textsc{pyGravSphere} density posteriors (Sec.  \ref{sec:data}) with no additional mass-dependent rescaling required. We stress that Eq.  \eqref{eq:SP} is not a new physical assumption relative to Eqs.  \eqref{eq:Schrod-std}--  \eqref{eq:Poisson-std}; it is the identical Klein--Gordon physics of Sec.  \ref{sec:action}, expressed in the field normalization adopted by our code.

\subsection{Madelung transformation and the quantum potential}

Writing $\Psi = \sqrt{\rho}\,\exp(iS/\hbar)$ (the Madelung transformation, Ref. \cite{Madelung1927}) and defining the velocity field $\mathbf{v}\equiv \nabla S/m$, the real and imaginary parts of Eq. \eqref{eq:SP} separate into a pair of hydrodynamic-like equations: a continuity equation, $\partial_t\rho + \nabla\cdot(\rho\mathbf{v}) = 0$, and a quantum Euler equation
\begin{equation}
    \partial_t \mathbf{v} + (\mathbf{v}\cdot\nabla)\mathbf{v} = -\nabla\Phi - \nabla Q ,
    \label{eq:eulerQ}
\end{equation}
where
\begin{equation}
    Q \equiv -\frac{\hbar^2}{2m^2}\,\frac{\nabla^2\sqrt{\rho}}{\sqrt{\rho}}
    \label{eq:Qpot}
\end{equation}
is the \emph{quantum potential} (or Bohm potential), the sole term distinguishing FDM dynamics from pressureless CDM at the fluid level. We emphasize that $Q$ depends on $\rho$ only through the logarithmic curvature of $\sqrt{\rho}$ (Eq. \ref{eq:Qform} below), and is therefore \emph{identical} whether $\rho$ is normalized as $|\Psi|^2$ (our convention, Eq. \ref{eq:SP}) or as $m|\psi_{\rm std}|^2$ (the standard convention, Eq. \ref{eq:Poisson-std}): a normalization choice for the wavefunction amplitude cannot, by itself, alter a physical density profile. The gravitational term $-\nabla\Phi$ in Eq. \eqref{eq:eulerQ} is likewise unaffected by this choice, since it descends from the $m\Phi\Psi$ term in Eq. \eqref{eq:SP}, which retains its coefficient of $m$ under the rescaling of Sec. \ref{sec:normalization}. The $\rho=|\Psi|^2$ convention therefore has no effect on any of the dynamical content of the equilibrium condition derived below; it is purely a bookkeeping choice, adopted here for direct consistency with our numerical implementation.

\subsection{Static, spherically symmetric equilibrium}

We are interested in the regime of an approximately static, spherically symmetric, curl-free ($\mathbf{v}\simeq0$, or more precisely a steady state in which the time-averaged bulk flow is negligible) DM distribution, appropriate for the relaxed inner regions of a dSph halo. Under these assumptions Eq. \eqref{eq:eulerQ} reduces to a condition of local hydrostatic balance,
\begin{equation}
    \nabla \Phi(r) = -\nabla Q(r) ,
    \label{eq:balance}
\end{equation}
i.e., the classical gravitational acceleration $g(r) \equiv d\Phi/dr = G M(<r)/r^2$ must be balanced, at every radius, by the gradient of the quantum pressure term sourced by the curvature of $\sqrt{\rho(r)}$. This is the equilibrium condition satisfied identically (by construction) by the ground-state SP soliton profile  \cite{Schive2014a,MarshPop2015,Chavanis2011a,ChavanisDelfini2011}, but Eq. \eqref{eq:balance} is more general: it can be evaluated for \emph{any} measured $\rho(r)$ and $M(<r)$, without assuming the density is in fact given by the analytic soliton solution.

For a spherically symmetric $\rho(r)$, Eq. \eqref{eq:Qpot} may be written compactly in terms of the logarithmic density slope. Defining $s(r) \equiv \ln\rho(r)$ and its log-radial derivative $a(r) \equiv \tfrac{1}{2}\,d\ln\rho/d\ln r$, the radial Laplacian of $\sqrt{\rho}=e^{s/2}$ gives
\begin{equation}
    Q(r) = -\frac{\hbar^2}{2m^2\,r^2}\left[ \frac{da}{d\ln r} + a(r) + a(r)^2 \right] .
    \label{eq:Qform}
\end{equation}
Equation \eqref{eq:Qform} depends only on the density profile's logarithmic derivatives, is straightforward to evaluate numerically by finite differencing $\ln\rho$ on a discrete radial grid, and requires no assumption about the functional form of $\rho(r)$. Substituting Eq. \eqref{eq:Qform} into the equilibrium condition \eqref{eq:balance} and rearranging for $m^2$ yields, at every radius where $dQ/dr$ and $g(r)$ are both well defined and of consistent sign,
\begin{equation}
    m^2(r) = -\hbar^2\,\frac{dQ_0/dr}{g(r)}
    \;\;\Longleftrightarrow\;\;
    \frac{1}{m^2} = -\frac{g(r)}{\hbar^2\,dQ_0/dr} ,
    \label{eq:localmass}
\end{equation}
where $Q_0(r) \equiv -\tfrac{1}{2r^2}\left[da/d\ln r + a + a^2\right]$ is the geometric (mass-independent) part of the quantum potential. Equation \eqref{eq:localmass} defines a \emph{local, radius-dependent} equilibrium mass estimator, $m_{\rm local}(r)$, which we use as a diagnostic (Sec. \ref{sec:results}) prior to combining information across radii in the global regression of Sec. \ref{sec:method}.

\subsection{Caveats of the static assumption}
\label{sec:caveats}

We emphasize that Eq. \eqref{eq:balance} assumes a strictly static, curl-free velocity field. Real dSph halos need not satisfy this exactly: residual bulk flows, non-zero angular momentum, tidal perturbation by the Milky Way potential, and departures from spherical symmetry can all source additional terms in the quantum Euler equation that are neglected here. The force-balance residual introduced in Sec. \ref{sec:robustness} is designed, in part, to flag galaxies or posterior realizations for which this assumption is a poor description of the data.

\section{Data}
\label{sec:data}

We use non-parametric density, $\rho(r)$, and enclosed dark-matter-mass, $M(<r)$, posterior chains for the nine Milky Way dSphs -- Carina, Canes Venatici~I (CVn~I), Draco, Fornax, Leo~I, Leo~II, Sculptor, Sextans, and Ursa Minor -- as reconstructed by \textsc{pyGravSphere}  \cite{pyGravSphere} (summarized in Sec. \ref{sec:pygs}), a non-parametric higher-order Jeans code that fits the spherical Jeans equation together with the fourth-order virial shape parameters to resolved line-of-sight stellar kinematics, thereby mitigating the well-known mass-velocity-anisotropy degeneracy that afflicts simpler single-moment Jeans analyses  \cite{ReadSteger2017,GonzalezMorales2017,Genina2020}. \textsc{GravSphere} has been validated against mock data spanning cored, cusped, and tidally stripped systems  \cite{ReadSteger2017,Genina2020} and applied to individual dSphs including Draco  \cite{ReadDraco2018}. For each galaxy we use $N_{\rm total}=50\,000$ posterior realizations of $\log_{10}\rho(r)$ and $\log_{10}M(<r)$ on a common radial grid spanning approximately $0.02$--$2\,\kpc$, log-spaced to match the resolution of the underlying binned stellar-kinematic data. The resolved line-of-sight velocity samples underlying these posteriors are drawn from the published kinematic datasets distributed with \textsc{pyGravSphere}: Carina, Fornax, Sculptor, and Sextans from Walker \textit{et al.} \cite{Walker2009}; Draco from Walker \textit{et al.} \cite{Walker2015}; Leo~I from Mateo \textit{et al.} \cite{Mateo2008}; Leo~II from Spencer \textit{et al.} \cite{Spencer2017}; Ursa~Minor from Spencer \textit{et al.} \cite{Spencer2018}; and CVn~I from Simon and Geha \cite{SimonGeha2007} and Martin \textit{et al.} \cite{Martin2007}.

\section{Mass modelling with \textsc{pyGravSphere}}
\label{sec:pygs}

The density and enclosed-mass posteriors used throughout this work (Sec. \ref{sec:data}) are
taken directly from \textsc{pyGravSphere}  \cite{pyGravSphere}, a \textsc{Python} implementation
of the \textsc{GravSphere} code of Refs. \cite{ReadSteger2017,ReadDraco2018,Genina2020}. For
completeness, and because the logarithmic curvature of $\rho(r)$ is the single quantity on which
our force-balance estimator depends (Eq. \ref{eq:Qform}), we summarize here the assumptions,
model components, and validation status of that pipeline. Nothing in this section is specific to
FDM: the kinematic modelling step is performed without any reference to the
Schr\"odinger--Poisson system, which is precisely what makes the equilibrium test of
Sec. \ref{sec:method} non-circular.

\subsection{Spherical Jeans equation and projection}
\label{sec:jeans}

Assuming spherical symmetry, dynamical equilibrium, and negligible rotation, the stellar tracer
population satisfies the spherical Jeans equation,
\begin{equation}
    \frac{1}{\nu(r)}\frac{d}{dr}\!\left[\nu(r)\,\sigma_r^2(r)\right]
    + \frac{2\beta(r)\,\sigma_r^2(r)}{r}
    = -\frac{G\,M(<r)}{r^2},
    \label{eq:app_jeans}
\end{equation}
where $\nu(r)$ is the three-dimensional tracer number density, $\sigma_r(r)$ the radial velocity
dispersion, $\beta(r) \equiv 1 - \sigma_\theta^2/\sigma_r^2$ the velocity anisotropy
($\beta=0$ isotropic, $\beta \to 1$ radial, $\beta<0$ tangential), and $M(<r)$ the enclosed mass.
Note that $M(<r)$ in Eq. \eqref{eq:app_jeans} is the same enclosed-mass profile that supplies the
classical gravitational acceleration $g(r) = GM(<r)/r^2$ in our equilibrium condition,
Eq. \eqref{eq:balance}. The observable is the projected line-of-sight dispersion at projected
radius $R$,
\begin{equation}
    \sigma_{\rm LOS}^2(R) = \frac{2}{\Sigma(R)}\int_R^\infty
    \left(1 - \beta\,\frac{R^2}{r^2}\right)
    \frac{\nu\,\sigma_r^2\,r\,dr}{\sqrt{r^2-R^2}},
    \label{eq:app_projdisp}
\end{equation}
with $\Sigma(R)$ the projected tracer surface density.

\subsection{Breaking the mass--anisotropy degeneracy: virial shape parameters}
\label{sec:vsp}

Equation \eqref{eq:app_projdisp} suffers from the well-known mass--anisotropy degeneracy: distinct
combinations of $\beta(r)$ and $M(<r)$ reproduce the same $\sigma_{\rm LOS}(R)$, so second moments
alone cannot separate a cored halo with radial orbits from a cuspy halo with tangential orbits.
\textsc{GravSphere} breaks this degeneracy using the two virial shape parameters (VSPs)
 \cite{ReadSteger2017}, which are radially integrated fourth moments of the line-of-sight velocity
distribution,
\begin{align}
    v_{s1} &= \frac{2}{5}\int_0^\infty G\,M(<r)\,\nu\,\bigl(5-2\beta\bigr)\sigma_r^2\,r\,dr \nonumber\\
           &= \int_0^\infty \Sigma\,\langle v_{\rm LOS}^4\rangle\,R\,dR ,
    \label{eq:app_vs1}\\
    v_{s2} &= \frac{4}{35}\int_0^\infty G\,M(<r)\,\nu\,\bigl(7-6\beta\bigr)\sigma_r^2\,r^3\,dr \nonumber\\
           &= \int_0^\infty \Sigma\,\langle v_{\rm LOS}^4\rangle\,R^3\,dR .
    \label{eq:app_vs2}
\end{align}
Because the right-hand sides are measured directly from the kurtosis of the observed velocity
distribution while the left-hand sides depend on $M(<r)$ and $\beta(r)$, including $v_{s1}$ and
$v_{s2}$ in the likelihood substantially reduces the allowed $(\rho,\beta)$ volume, in particular
at small radii where our fitting window is most constraining.

\subsection{Model components}
\label{sec:components}

\paragraph{Tracer density.} The projected stellar light profile of each \dsph\ is fitted with a sum
of three Plummer spheres,
\begin{align}
    \Sigma(R) &= \sum_{k=1}^{3}\frac{L_k}{\pi a_k^2}
    \left[1+\left(\frac{R}{a_k}\right)^{\!2}\right]^{-2}, \label{eq:app_plummer}\\
    \nu(r) &= \sum_{k=1}^{3}\frac{3L_k}{4\pi a_k^3}
    \left[1+\left(\frac{r}{a_k}\right)^{\!2}\right]^{-5/2},
\end{align}
(where $L_k$ and $a_k$ are the total luminosity and plummer scale radius of $k_{th}$ component of the plummer model respectively) which is enough to describe the observed surface brightness without imposing a
one-parameter shape, and which admits the analytic deprojection quoted above.

\paragraph{Dark-matter density.} The DM density is \emph{not} assumed to follow any analytic halo
form. \textsc{GravSphere} instead models $\rho(r)$ as a broken power law defined on a set of fixed
radial bins tied to the projected half-light radius $R_{1/2}$,
\begin{equation}
\begin{split}
    \rho(r) &= \rho_n \left(\frac{r}{r_n}\right)^{-\gamma_n},
    \qquad r_n < r < r_{n+1}, \\
    r_n/R_{1/2} &\in \{0.25,\,0.5,\,1,\,2,\,4\},
\end{split}
    \label{eq:app_bpl}
\end{equation}
with the normalizations matched across bins so that $\rho(r)$ is continuous, and with the bin
slopes $\gamma_n$ (and an overall normalization) treated as free parameters. This is the sense in
which the $\rho(r)$ and $M(<r)$ posteriors used in this paper are \emph{non-parametric}: the inner
logarithmic slope is inferred from the data rather than fixed by a cusped or cored template, and no
soliton, NFW, or Zhao  \cite{Zhao1996} functional form is imposed. This property is essential for our
analysis, since the quantum-potential term $Q_0(r)$ of Eq. \eqref{eq:Qform} depends on the second
logarithmic derivative of $\rho$; had a soliton-like profile been assumed at the modelling stage, the
force-balance test of Sec. \ref{sec:method} would be circular by construction. It also means, as
discussed in Sec. \ref{sec:results}, that there is no reason for the recovered curvature of
$\rho(r)$ to match that of an analytic ground-state soliton at every radius. We note a limitation of this construction: because $\rho(r)$ is piecewise power-law between the fixed bin edges, $da/d\ln r$ vanishes inside each bin, so the curvature entering $Q_0(r)$ comes from the slope changes at the bin edges (as smoothed by finite differencing on the output grid), and otherwise $Q_0\simeq\gamma(2-\gamma)/(8r^2)$ simply follows the local slope. ``Non-parametric'' therefore means that no soliton, NFW or other analytic template is imposed, not that the second logarithmic derivative of $\rho$ is independently resolved by the data.

\paragraph{Velocity anisotropy.} The anisotropy profile is parametrized in the symmetrized form
\begin{equation}
    \tilde{\beta}(r) \equiv \frac{\sigma_r^2-\sigma_t^2}{\sigma_r^2+\sigma_t^2}
    = \tilde{\beta}_0 + \bigl(\tilde{\beta}_\infty - \tilde{\beta}_0\bigr)
      \frac{1}{1+\left(r_0/r\right)^{n}},
    \label{eq:app_beta}
\end{equation}
which interpolates smoothly between an inner value $\tilde{\beta}_0$ and an outer value
$\tilde{\beta}_\infty$ across a transition radius $r_0$ with sharpness $n$, and which remains bounded
for strongly tangential orbits.

\subsection{Posterior sampling}
\label{sec:mcmc}

The free parameters of Eqs. \eqref{eq:app_bpl}-- \eqref{eq:app_beta}, together with the tracer and
nuisance parameters, are sampled with the \texttt{emcee} affine-invariant ensemble sampler
 \cite{ForemanMackey2013}, using a likelihood built from the binned $\sigma_{\rm LOS}(R)$ profile and
the two VSPs. The output is a Markov chain of density and enclosed-mass realizations
$\{\rho_j(r),\,M_j(<r)\}$ on a common radial grid; the posterior median and the $68\%$/$95\%$
credible bands of these chains are what is plotted in Fig. \ref{fig:rho}, and each
individual realization $j$ is propagated independently through the force-balance regression of
Eq. \eqref{eq:regression}. Our quoted per-galaxy $m_{21}$ posteriors therefore inherit the full
kinematic uncertainty of the \textsc{pyGravSphere} chains, rather than only the uncertainty on a
median profile.

\subsection{Validation on mock data}
\label{sec:validation}

Read \textit{et al.} \cite{Read2021} compared \textsc{GravSphere} with three independent mass-orbit modelling methods
(DiscreteJAM, MAMPOSSt, and the \textsc{Agama} distribution-function fitter) on the Gaia Challenge
suite of spherical mocks, which span cored and cusped halos, isotropic, Osipkov--Merritt, and
tangential anisotropy, and tracer samples of $10^3$--$10^4$ stars with $2\,\mathrm{km\,s^{-1}}$
velocity errors. For isotropic and tangentially anisotropic mocks, all methods recover $\rho(r)$ and
$\beta(r)$ within their $95\%$ credible intervals over $0.25 \lesssim R/R_{1/2} \lesssim 4$. The
radially anisotropic mocks are the most demanding case, and with line-of-sight data alone only
methods that use information about the shape of the velocity distribution recover both profiles
without bias; \textsc{GravSphere} does so through the VSPs, whereas second-moment-only methods are
systematically biased towards tangential anisotropy and underestimate $\rho(r)$ beyond $R_{1/2}$.
Further tests on cosmological and tidally stripped dwarfs are presented in Genina \textit{et al.} \cite{Genina2020} and
Collins \textit{et al.} \cite{Collins2021}.

Two consequences for the present work follow. First, the radial range over which the reconstruction has been validated is
$0.25\,R_{1/2} \lesssim r \lesssim 4\,R_{1/2}$. The fiducial fitting window $(r_{\min},r_{\max}) = (0.05,\,1.0)\,\kpc$ was chosen to avoid the innermost and outermost grid points, where the finite-difference noise in $Q_0(r)$ (Sec. \ref{sec:method}) and the weak kinematic constraint on $M(<r)$ are most severe, and it is not tied to $R_{1/2}$. Since the half-light radii of the classical dSphs are a few hundred pc, its lower edge lies below $0.25\,R_{1/2}$ for most of the sample, so the inner part of the window probes a regime in which the reconstruction has not been validated on mock data. The window-sensitivity test of Fig. \ref{fig:sensitivity} quantifies the consequences. Second, the regime expected for an FDM halo mildly radial anisotropy
in the granular envelope outside the coherent core is precisely the regime in which the VSPs are
required, which is the principal reason we use \textsc{pyGravSphere} posteriors rather than
single moment Jeans fits. We note finally that Zimmermann \textit{et al.} \cite{Zimmermann2025}, whose Leo~II bound we compare
against in Sec. \ref{sec:discussion}, adopt the coreNFWtides parametrization
 \cite{ReadSteger2017} rather than the broken power law of Eq. \eqref{eq:app_bpl}; assessing whether
this choice shifts the recovered $m_{21}$ at the level relevant to our constraints is left to future
work.

\section{Method}
\label{sec:method}

\subsection{Per-posterior force-balance regression}

For each posterior realization $j = 1,\dots,N_{\rm post}$ of a given galaxy, we compute the quantum-potential term $Q_{0,j}(r)$ and its radial derivative $Q_{0,j}'(r) \equiv dQ_{0,j}/dr$ via finite differences of $\ln\rho_j(r)$ on the native radial grid (Eq. \ref{eq:Qform}), and the classical gravitational acceleration $g_j(r) = G\,M_j(<r)/r^2$ from the corresponding enclosed-mass posterior. Rather than solving Eq. \eqref{eq:localmass} independently at every radius which is numerically unstable wherever $g_j(r)$ or $Q_{0,j}'(r)$ pass through zero or become dominated by finite-difference noise we instead solve for a single, radius independent $\alpha_j \equiv \hbar^2/m_j^2$ per posterior draw via an unweighted least-squares minimization of the fractional force-balance residual over a restricted radial window $[r_{\min},r_{\max}]$:
\begin{equation}
    \alpha_j = \operatorname*{arg\,min}_{\alpha} \sum_{r_i \in [r_{\min},r_{\max}]} \left(\frac{g_j(r_i) + \alpha\, Q_{0,j}'(r_i)}{g_j(r_i)}\right)^{\!2} ,
    \label{eq:regression}
\end{equation}
which admits the closed-form solution $\alpha_j = -\sum_i x_i/\sum_i x_i^2$ with $x_i \equiv Q_{0,j}'(r_i)/g_j(r_i)$. Radial points are included in the sum only if both $g_j(r_i)>0$ and $Q_{0,j}'(r_i)<0$ (the sign combination required for a physically sensible, positive $\alpha_j$) and if $r_i$ lies within a fiducial window $r_{\min}=0.05\,\kpc$, $r_{\max}=1.0\,\kpc$, chosen to exclude the innermost bins (where finite-differencing $\ln\rho$ is most susceptible to shot noise in the GravSphere posterior) and the outermost bins (where the posterior on $M(<r)$ is least constrained by the stellar kinematic data). We assess the sensitivity of our results to this choice explicitly in Sec. \ref{sec:robustness}. Draws for which fewer than twenty five radial points satisfy these criteria, or for which the resulting $\alpha_j \le 0$, are discarded as non-converged.

The implied boson mass for posterior draw $j$ follows directly from the definition of $\alpha_j$,
\begin{equation}
    m_j = \frac{\hbar}{\sqrt{\alpha_j}} ,
    \qquad
    m_{21} \equiv \frac{m_j c^2}{10^{-21}\,\mathrm{eV}} ,
\end{equation}
and we report, for each galaxy, the resulting distribution $\{m_{21}^{(j)}\}$ over all converged posterior draws.

\subsection{Fit-quality diagnostic}

For each converged draw we additionally record the force-balance residual sum,
\begin{equation}
    S_j \equiv \sum_{r_i} \left(\frac{g_j(r_i) + \alpha_j\, Q_{0,j}'(r_i)}{g_j(r_i)}\right)^{\!2} ,
    \label{eq:residualsum}
\end{equation}
and the number of radial points entering the fit, $n_{{\rm used},j}$. We stress that $S_j$ is a raw, unweighted sum of squared fractional residuals it is \emph{not} normalized by the GravSphere posterior uncertainties on $\rho(r)$ or $M(<r)$, nor by the number of degrees of freedom, and should therefore not be interpreted as a conventional $\chi^2$ statistic in the sense of following a $\chi^2$ distribution. To allow meaningful comparison across posterior draws and galaxies with different $n_{{\rm used},j}$, we report the \emph{reduced residual}, $S_j/n_{{\rm used},j}$, throughout.

\subsection{Numerical implementation}
\label{sec:numerics}

The estimator is evaluated draw by draw on the full set of \textsc{pyGravSphere} chains. For each galaxy we start from the $N_{\rm total}=50\,000$ posterior realizations of $\log_{10}\rho_j(r)$ and $\log_{10}M_j(<r)$ tabulated on a common log-spaced radial grid spanning $\approx0.02$--$2\,\kpc$ (Sec. \ref{sec:data}). For each draw $j$:
\begin{enumerate}
\item Convert the stored logarithmic quantities to the linear density $\rho_j(r)$ and enclosed mass $M_j(<r)$.
\item Compute the logarithmic slope $a_j(r)=\tfrac12\,d\ln\rho_j/d\ln r$ and its derivative $da_j/d\ln r$ by finite differences on the native grid, evaluate $Q_{0,j}(r)$ from Eq. \eqref{eq:Qform}, and obtain $Q_{0,j}'(r)=dQ_{0,j}/dr$ by finite differences.
\item Compute the classical acceleration $g_j(r)=GM_j(<r)/r^2$.
\item Select the radial points with $r_{\min}\le r_i\le r_{\max}$, $g_j(r_i)>0$, and $Q_{0,j}'(r_i)<0$. If fewer than twenty five points survive, the draw is discarded as non-converged.
\item Form $x_i=Q_{0,j}'(r_i)/g_j(r_i)$ and evaluate the closed-form least-squares solution $\alpha_j=-\sum_i x_i/\sum_i x_i^2$ of Eq. \eqref{eq:regression}; no iterative optimizer is needed. Discard the draw if $\alpha_j\le0$.
\item Convert to a boson mass, $m_j=\hbar/\sqrt{\alpha_j}$ and $m_{21}^{(j)}=m_jc^2/(10^{-21}\,\mathrm{eV})$, and record the residual sum $S_j$ and the number of points $n_{{\rm used},j}$ (Eq. \ref{eq:residualsum}).
\end{enumerate}
The per-galaxy summary statistics are the median and the 16th and 84th percentiles of $\{m_{21}^{(j)}\}$ over converged draws, together with the convergence fraction $N_{\rm valid}/N_{\rm total}$ (Table \ref{tab:summary}); the inverse-variance-weighted effective equilibrium mass follows from Eqs. \eqref{eq:weightedmean}-- \eqref{eq:weightedsigma}. Because each draw $j$ is propagated independently through steps~1--6, the per-galaxy posteriors inherit the full kinematic uncertainty of the \textsc{pyGravSphere} chains rather than only the uncertainty on a median profile. The second logarithmic derivative entering $Q_0$ makes it the noisiest quantity in the pipeline (Fig. \ref{fig:q0}), which is why the innermost and outermost grid points are excluded by the fiducial window.

\section{Results}
\label{sec:results}

\subsection{Density, acceleration, and quantum-pressure profiles}

\begin{figure*}[tp]
    \centering
    \includegraphics[width=0.92\textwidth]{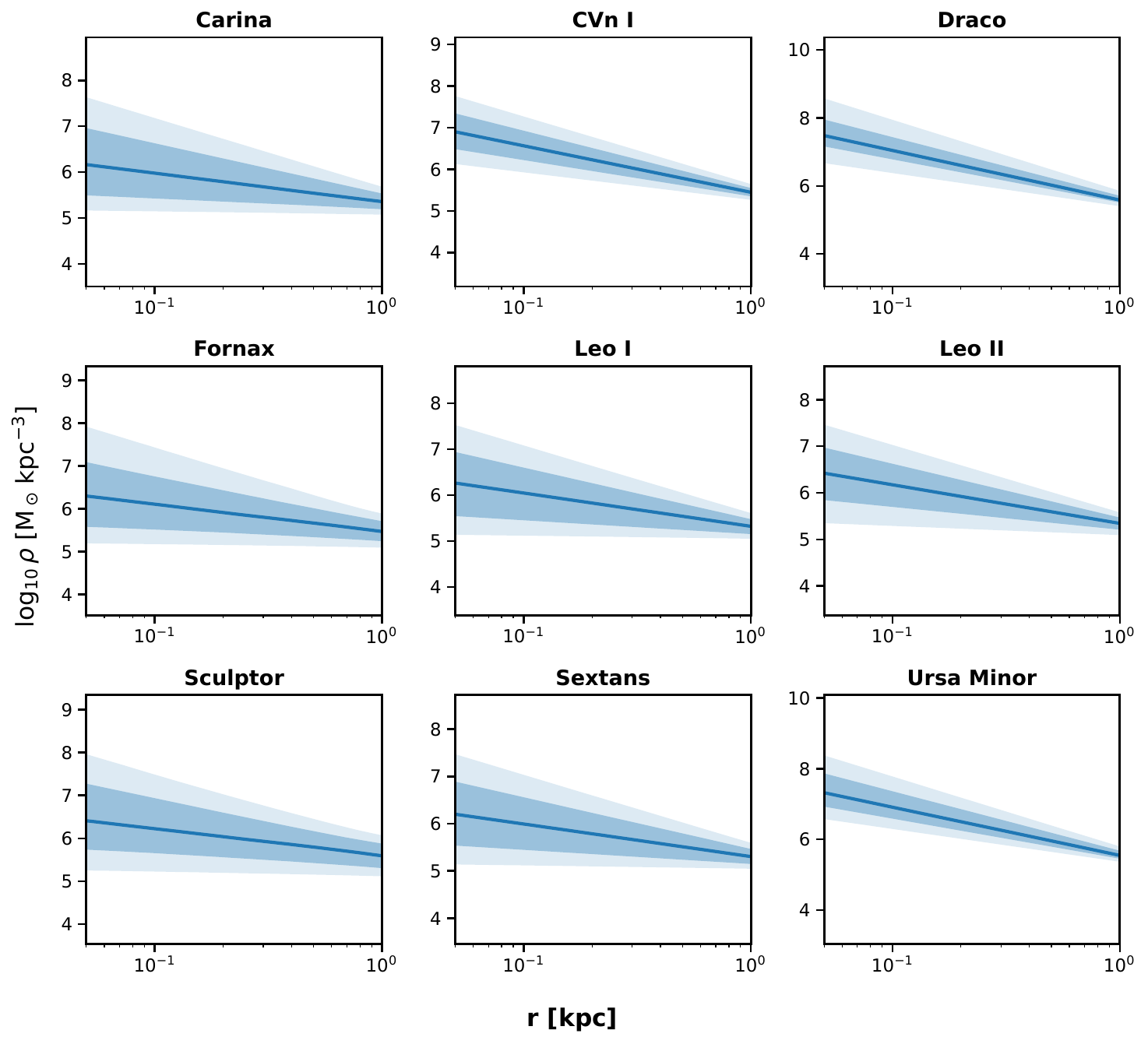}
    \caption{Dark-matter density profiles for all nine Milky Way dSphs analyzed in this work (one panel per galaxy, arranged in a $3\times3$ grid; galaxy name given in each panel title). The horizontal axis is galactocentric radius $r$ on a logarithmic scale, showing the portion $\approx0.05$--$1\,\kpc$ of the common \textsc{pyGravSphere} grid ($0.02$--$2\,\kpc$); the vertical axis is the logarithm of the dark-matter density, $\log_{10}[\rho(r)/(\msun\,\mathrm{kpc}^{-3})]$. The solid curve in each panel is the posterior-median profile across all retained \textsc{pyGravSphere} Markov-chain realizations for that galaxy (Sec. \ref{sec:data}), obtained by fitting the spherical Jeans equation together with higher-order virial shape parameters to the resolved line-of-sight stellar kinematics  \cite{ReadSteger2017}. The shaded bands show the $68\%$ (dark) and $95\%$ (light) credible intervals at each radius, i.e.\ the $16$th--$84$th and $2.5$th--$97.5$th percentile envelopes of $\log_{10}\rho(r)$ across posterior draws; these bands are pointwise in $r$ and are not simultaneous confidence regions. This figure is purely a data-quality and consistency check: it does not itself enter the force-balance regression of Sec. \ref{sec:method}, but the logarithmic curvature of these profiles is precisely the quantity from which the quantum-potential term $Q_0(r)$ (Fig. \ref{fig:q0}) is computed via Eq. \eqref{eq:Qform}.}
    \label{fig:rho}
\end{figure*}

\begin{figure*}[tp]
    \centering
    \includegraphics[width=0.92\textwidth]{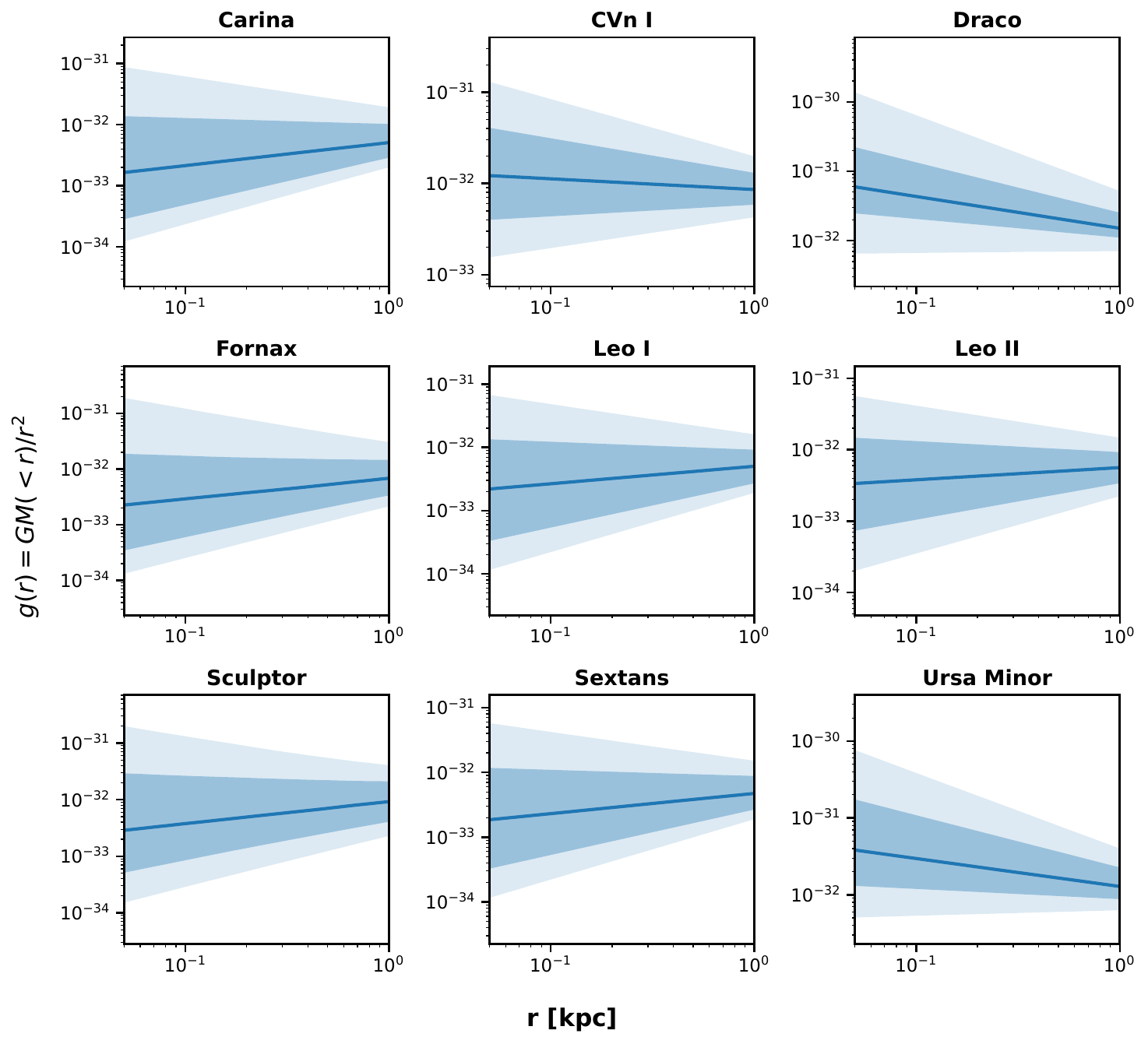}
    \caption{Classical Newtonian gravitational acceleration $g(r) = GM(<r)/r^2$ for all nine dSphs, computed pointwise from the enclosed-mass posterior for each realization $j$ and each radius on the native grid. Panel layout, axis conventions (log--log), and credible-band definitions follow Fig. \ref{fig:rho} and Fig.  \ref{fig:q0}. This is the ``classical gravity'' side of the local hydrostatic-equilibrium condition, Eq. \eqref{eq:balance}; the force-balance regression of Sec. \ref{sec:method} solves for the single boson mass $m_j$ per posterior draw that best equates $g_j(r)$ to the (mass-dependent) quantum-pressure gradient $-\alpha_j\,Q_{0,j}'(r)$ over the fitting window $[r_{\min},r_{\max}]$.}
    \label{fig:g}
\end{figure*}

\begin{figure*}[tp]
    \centering
    \includegraphics[width=0.92\textwidth]{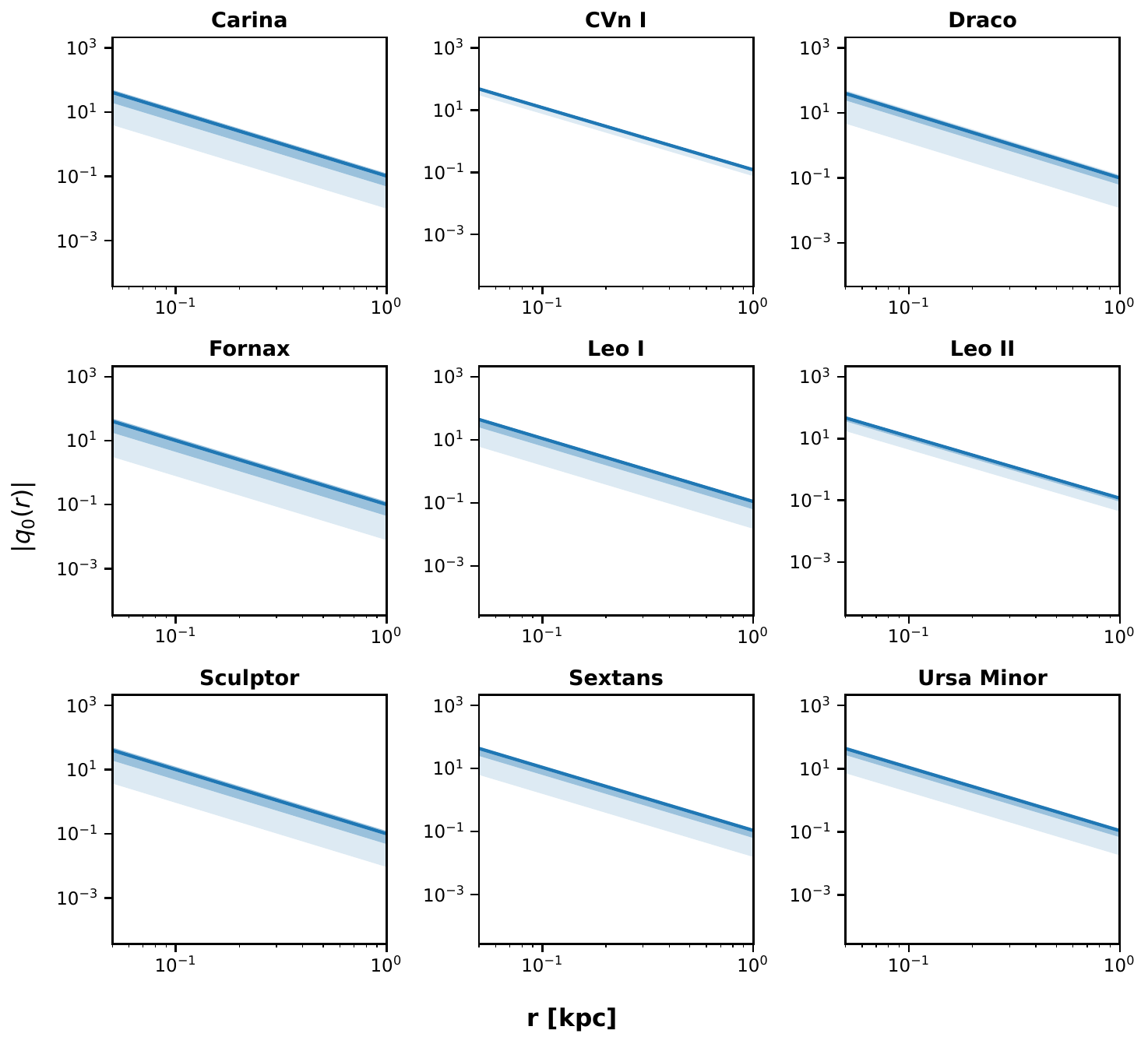}
    \caption{Geometric (mass-independent) quantum-potential term $|Q_0(r)|$ for all nine dSphs, in the same panel layout as Fig. \ref{fig:rho}. Here $Q_0(r) \equiv -\tfrac{1}{2r^2}\left[da/d\ln r + a(r) + a(r)^2\right]$, with $a(r)\equiv\tfrac{1}{2}\,d\ln\rho/d\ln r$ the logarithmic density slope (Eq. \ref{eq:Qform}); $Q_0$ carries the full radial and posterior realization dependence of the quantum potential $Q(r) =(\hbar^2/m^2)\,Q_0(r)$ but, unlike $Q(r)$ itself, is independent of the (unknown) boson mass $m$, since it is computed purely from finite differences of $\ln\rho_j(r)$ on the native \textsc{pyGravSphere} radial grid for each posterior draw $j$. The horizontal axis is galactocentric radius $r$ (log scale); the vertical axis shows $|Q_0(r)|$ on a logarithmic scale, with the posterior-median curve and $68\%/95\%$ credible bands defined as in Fig. \ref{fig:rho}. Because $Q_0(r)$ involves a second logarithmic derivative of the density, it is the noisiest of the quantities entering our analysis; sharp features or large credible bands in this figure directly propagate into the local mass estimator $m_{\rm local}(r)$ of Fig. \ref{fig:mlocal} and motivate the restricted radial fitting window used in Eq. \eqref{eq:regression}.}
    \label{fig:q0}
\end{figure*}

Figure \ref{fig:rho} shows the posterior-median $\rho(r)$ profile, with $68\%$ and $95\%$ credible bands, for all nine dSphs reconstructed by \textsc{pyGravSphere}. Figure \ref{fig:g} shows the classical gravitational acceleration $g(r)$ computed from the enclosed mass, and Fig. \ref{fig:q0} shows the geometric, mass-independent quantum-potential term $Q_0(r)$ (Eq. \ref{eq:Qform}). The dimensionless ratio $x(r)=Q_0'(r)/g(r)$ that enters the regression of Eq. \eqref{eq:regression} follows directly from these two quantities. Together these figures document every ingredient entering the force-balance regression of Sec. \ref{sec:method}.

\subsection{Local equilibrium mass diagnostic}

\begin{figure*}[tp]
    \centering
    \includegraphics[width=0.92\textwidth]{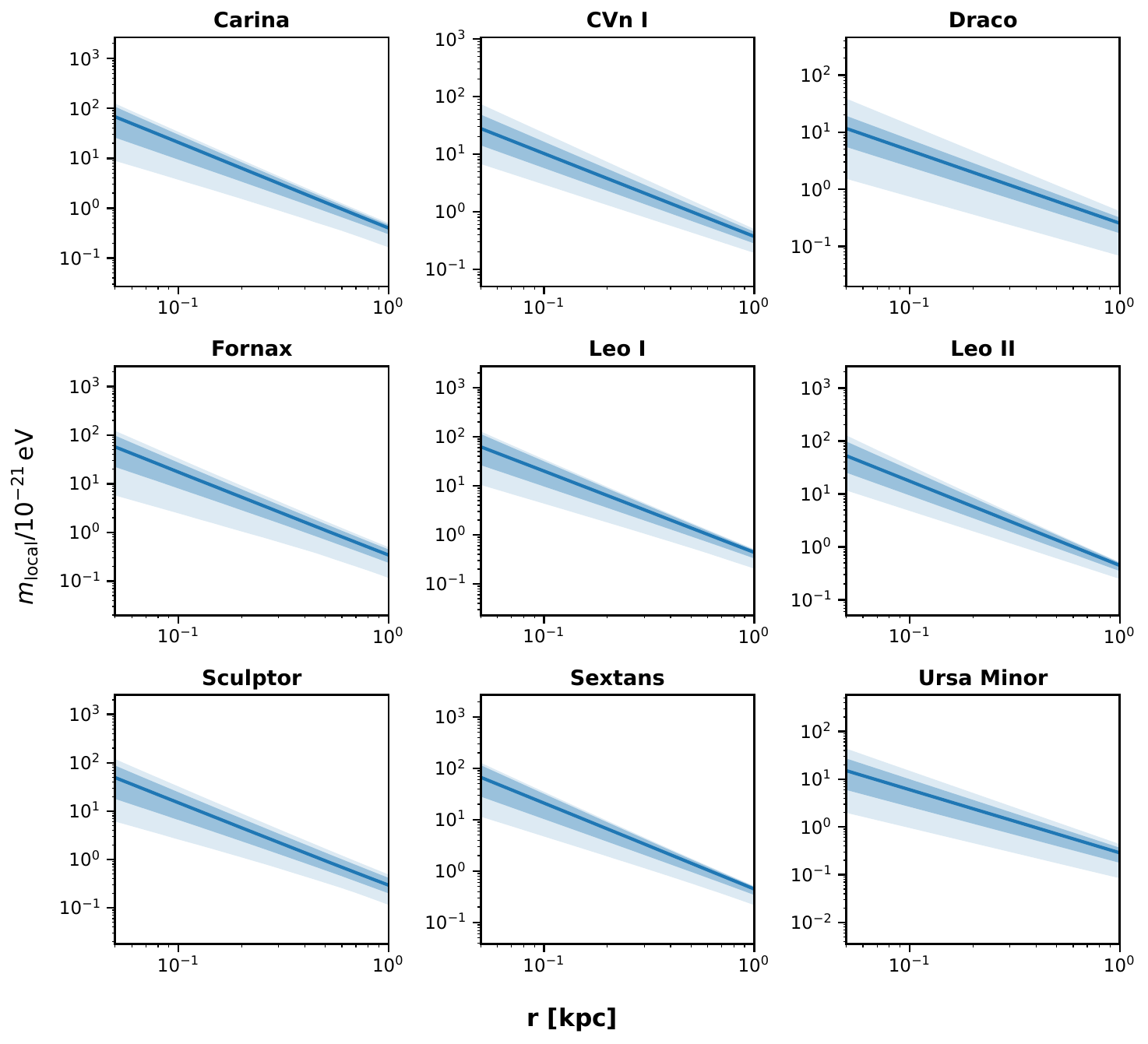}
    \caption{Local, single-radius equilibrium boson-mass estimator $m_{\rm local}(r)$ (in units of $m_{21}$), Eq. \eqref{eq:localmass}, for all nine dSphs, plotted only at radii where $g(r)>0$ and $Q_0'(r)<0$ (the sign combination required for a physically sensible, positive squared mass). Unlike the global regression of Eq. \eqref{eq:regression}, this estimator solves the equilibrium condition independently at each radius, without combining information across $r$; it is a diagnostic, not a summary statistic. The horizontal axis is galactocentric radius $r$ (log scale); the vertical axis is $m_{\rm local}(r)$ (log scale), with the posterior-median curve and $68\%/95\%$ credible bands defined as in Fig. \ref{fig:rho}. For every galaxy, $m_{\rm local}(r)$ declines smoothly and monotonically by roughly one and a half to two and a half orders of magnitude across the plotted range, rather than remaining flat; this is close to the generic expectation for a non-parametric \textsc{pyGravSphere} density profile that does not share the specific logarithmic curvature of the analytic SP soliton solution at every radius (see the discussion above and Sec. \ref{sec:discussion} for physical interpretation). The absence of a flat plateau means that the single, radius-independent $\alpha_j = \hbar^2/m_j^2$ fitted in Eq. \eqref{eq:regression} should be read as an effective, fitting-window-dependent equilibrium mass rather than a value independently confirmed across the full radial range; this motivates the restricted fitting window used in Eq. \eqref{eq:regression} and the explicit window-sensitivity test of Fig. \ref{fig:sensitivity}.}
    \label{fig:mlocal}
\end{figure*}

Before combining information across radii, we examine the purely local, single-radius estimator $m_{\rm local}(r)$ of Eq. \eqref{eq:localmass} (Figure \ref{fig:mlocal}). In principle, a radially flat $m_{\rm local}(r)$ over an extended range would indicate that the assumption of a single, radius-independent boson mass satisfying local equilibrium is an accurate description of the data over that range. In practice, we find instead that $m_{\rm local}(r)$ declines smoothly and monotonically by roughly one and a half to two and a half orders of magnitude across the plotted range ($r\approx0.05$--$1\,\kpc$) for every galaxy in our sample. We discuss the origin and implications of this trend below, since it directly affects how the global regression of Sec. \ref{sec:method} should be interpreted.

The decline is, in fact, close to the generic expectation for a non-parametric profile that is not itself a soliton. From Eq. \eqref{eq:localmass}, $m_{\rm local}^2(r) = -\hbar^2\,Q_0'(r)/g(r)$ is radius-independent only if $\rho(r)$ has exactly the logarithmic curvature of the analytic SP ground-state soliton at every radius; this is the defining property of the soliton solution and is not guaranteed for an arbitrary reconstructed profile. For a locally power-law density, $\rho(r)\propto r^{-\gamma}$ with $0<\gamma<2$, Eq. \eqref{eq:Qform} gives $Q_0(r)=\gamma(2-\gamma)/(8r^2)$ and hence $Q_0'(r)\propto r^{-3}$, while $g(r)=GM(<r)/r^2$ depends on $\gamma$ through the enclosed mass and is roughly flat in the cored inner regions. Equation \eqref{eq:localmass} then reads $m_{\rm local}^2=\hbar^2\gamma(2-\gamma)/\bigl(4GM(<r)\,r\bigr)$, i.e.\ $m_{\rm local}\simeq\hbar\sqrt{\gamma(2-\gamma)}/(2\,r\,v_c)$ with $v_c^2=GM(<r)/r$: the local estimator expresses the condition that the de~Broglie wavelength $\hbar/(m v_c)$ is comparable to $r$, and it falls with radius as $(g\,r^3)^{-1/2}\sim r^{-3/2}$ for roughly constant $g$. This is sufficient to produce a decline of $m_{\rm local}$ by roughly two orders of magnitude across about a decade in radius, comparable to what is seen in Fig. \ref{fig:mlocal}, without requiring any pathology in the data or the estimator.

We identify three, non-exclusive, physical contributions to this trend. First, and most directly, \textsc{pyGravSphere} reconstructs $\rho(r)$ and $M(<r)$ purely from the Jeans/virial-shape fit to the resolved stellar kinematics, with no reference to the FDM soliton solution; there is accordingly no reason for its recovered logarithmic curvature to match the soliton profile at all radii, and the observed decline can largely be read as a direct measurement of how much the reconstructed profile shape departs from a soliton with a single, fixed mass. Second, cosmological FDM simulations  \cite{Schive2014a,Veltmaat2018} find that the coherent, static soliton core persists only out to a soliton radius of order the de~Broglie scale; beyond this radius the halo transitions to a granular, wave-interference-supported envelope that is not, at any instant, in the static, curl-free equilibrium assumed in Sec. \ref{sec:theory}. If our fitting window extends beyond the true soliton radius of a given galaxy, the outer points included in the regression are probing envelope structure rather than core structure, which would bias the fitted $m_j$ away from the core-equilibrium value and is a plausible contributor to the tension with soliton-fit-based constraints discussed in Sec. \ref{sec:discussion}. Third, residual bulk flows, non-sphericity, and Milky-Way tidal perturbation -- the departures from the idealized assumptions of Sec. \ref{sec:theory} discussed in Sec. \ref{sec:caveats} -- are expected to grow in relative importance toward the outskirts of the halo, further degrading the equilibrium assumption at larger $r$.

Because $m_{\rm local}(r)$ is not flat, Eq. \eqref{eq:regression} should be understood as fitting an \emph{effective}, window-dependent
 equilibrium mass rather than recovering a single physical mass that is independently confirmed at each radius. This motivates both the restricted fitting window already adopted in Eq. \eqref{eq:regression} and the explicit window-sensitivity test of Sec. \ref{sec:robustness}, and it is the primary reason we report the global regression as a diagnostic equilibrium-mass estimate rather than as a direct measurement of the fundamental boson mass.

A further consequence concerns how the regression weights the radii. Since $x_i=Q_{0,j}'(r_i)/g_j(r_i)=-m_{{\rm local},j}^2(r_i)/\hbar^2$ varies by several orders of magnitude across the window, $\sum_i x_i^2$ is dominated by the innermost points, and Eq. \eqref{eq:regression} returns a value close to $m_{\rm local}$ evaluated near $r_{\min}$ rather than an average over the window. The strong window dependence found in Sec. \ref{sec:robustness}, where the medians of Fig. \ref{fig:sensitivity} scale roughly as $r_{\min}^{-1.3}$ to $r_{\min}^{-1.9}$, similar to the local slope of $m_{\rm local}$ in Fig. \ref{fig:mlocal}, is the expected consequence. The estimate should accordingly be read as the equilibrium mass of the reconstructed profile at the inner edge of the fitting window.

\subsection{Global per-galaxy boson-mass posteriors}

\begin{figure*}[tp]
    \centering
    \includegraphics[width=0.92\textwidth]{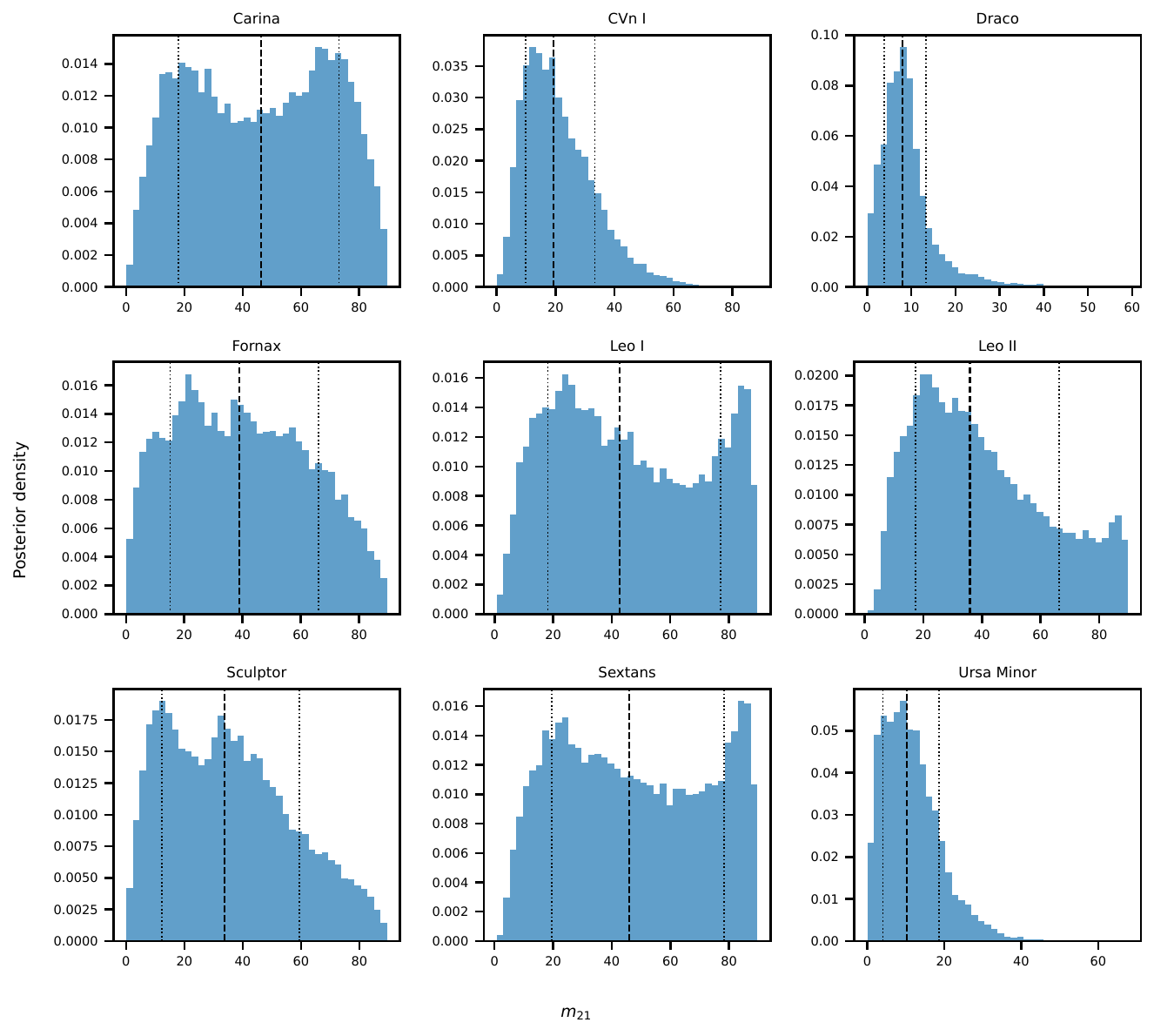}
    \caption{Posterior histograms of the implied boson mass $m_{21}\equiv m/10^{-21}\,\mathrm{eV}$ from the global per-posterior-draw force-balance regression (Eq. \ref{eq:regression}), for all nine dSphs, arranged in the same $3\times3$ grid as the preceding figures. Each histogram is built from the full set of converged \textsc{pyGravSphere} posterior draws for that galaxy (i.e.\ those satisfying $n_{\rm used}\ge25$ and $\alpha_j>0$; see Sec. \ref{sec:method} and the convergence fractions of Fig. \ref{fig:convergence} and Table \ref{tab:summary}). The horizontal axis is $m_{21}^{(j)}$, the boson mass implied by posterior draw $j$; the vertical axis is the probability density per unit $m_{21}$. In each panel the dashed line marks the posterior median and the two dotted lines mark the $16$th and $84$th percentiles reported numerically in Table \ref{tab:summary}. Galaxies with a broader or more skewed histogram (e.g.\ a long tail to high $m_{21}$) should be interpreted together with the residual-versus-mass diagnostic of Fig. \ref{fig:residual} (Sec. \ref{sec:robustness}), since the high-$m_{21}$ tail of each galaxy's distribution is preferentially populated by draws with larger residuals.}
    \label{fig:m22hist}
\end{figure*}

\begin{figure}[t]
    \centering
    \includegraphics[width=\columnwidth]{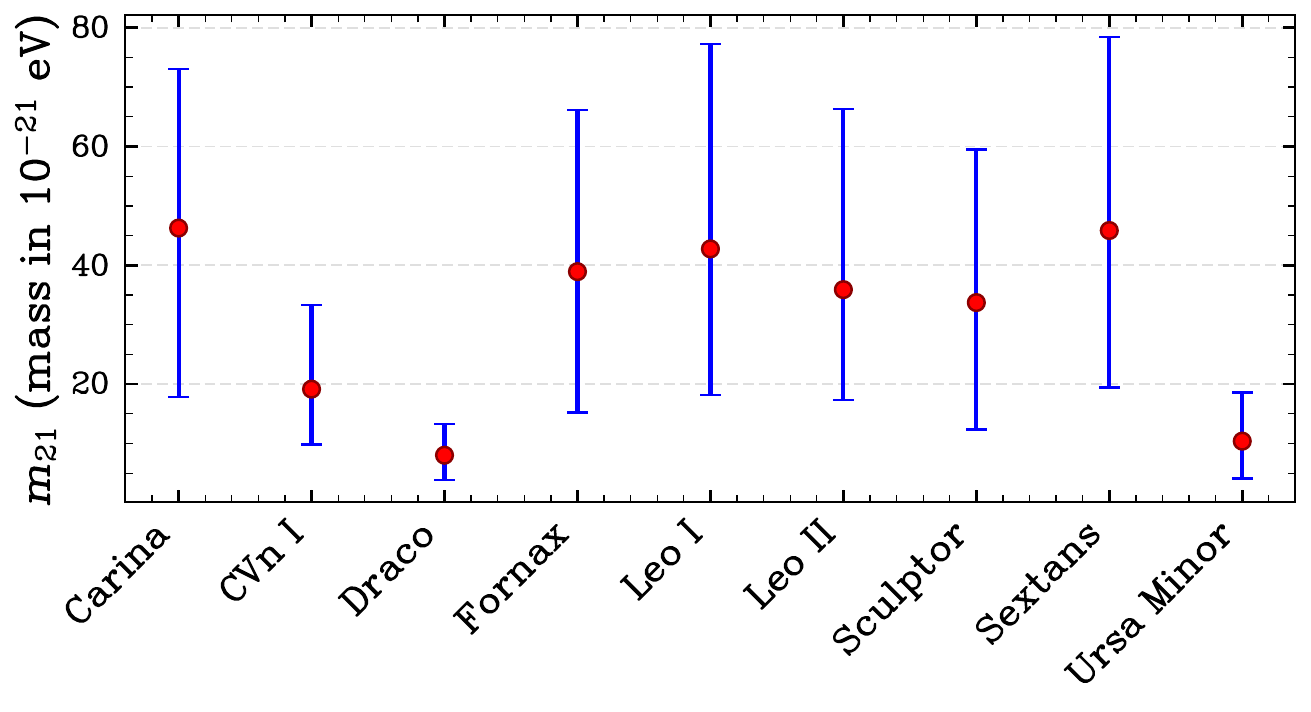}
    \caption{Summary of the boson mass $m_{21}\equiv m/10^{-21}\,\mathrm{eV}$ required for local hydrostatic equilibrium, one point per galaxy, for all nine dSphs analyzed in this work (galaxy name on the horizontal axis, ordered as in Table \ref{tab:summary}). Each point marks the posterior median of $m_{21}^{(j)}$ from the global force-balance regression (the same distributions histogrammed individually in Fig. \ref{fig:m22hist}), and the vertical error bar spans the asymmetric $68\%$ interval defined by the $16$th and $84$th percentiles of that posterior (i.e.\ the ``$+1\sigma$''/``$-1\sigma$'' columns of Table \ref{tab:summary}). This figure is the single-panel, cross-galaxy condensation of the nine individual posteriors in Fig. \ref{fig:m22hist} and provides the most direct visual comparison of the equilibrium-mass constraint across the sample; non-overlapping error bars indicate differences between the inferred equilibrium masses of individual galaxies under this method, which we discuss further in Sec. \ref{sec:discussion}.}
    \label{fig:summary}
\end{figure}

Figure \ref{fig:m22hist} shows the resulting $m_{21}$ posterior histograms for all nine dSphs, and Figure \ref{fig:summary} and Table \ref{tab:summary} summarize the per-galaxy median and $68\%$ interval, together with the fraction of posterior draws that satisfied our convergence criteria, $N_{\rm valid}/N_{\rm total}$ (see Sec. \ref{sec:robustness}). Throughout, the $68\%$ interval of $m_{21}$ denotes the $16$th--$84$th percentile range of the implied masses $m_{21}^{(j)}$ across the \textsc{pyGravSphere} posterior draws. Because each $m_{21}^{(j)}$ is a deterministic function of one draw of $(\rho,M)$ rather than the outcome of a likelihood-based inference on $m$, this range propagates the kinematic uncertainty of the reconstruction and is not a Bayesian credible interval for the boson mass.

\begin{table}[t]
\caption{Per-galaxy summary of the boson mass $m_{21}\equiv m/10^{-21}\,\mathrm{eV}$ required for local hydrostatic equilibrium, from the global force-balance regression of Eq. \eqref{eq:regression} applied to every retained \textsc{pyGravSphere} posterior draw. $m_{21}$ is the posterior median; $+1\sigma$ and $-1\sigma$ are the half-widths of the $68\%$ interval (84th percentile minus median, and median minus 16th percentile), so the interval is $m_{21}=\text{median}\,^{+1\sigma}_{-1\sigma}$. $N_{\rm valid}$ is the number of draws for which the regression converged (at least twenty five radial points with $g_j>0$ and $Q_{0,j}'<0$ in the window, and $\alpha_j>0$; Sec. \ref{sec:method}) and $N_{\rm total}=50\,000$ for every galaxy. All entries use the fiducial window $r_{\rm min}=0.05\,\kpc$, $r_{\rm max}=1.0\,\kpc$; the sensitivity to this choice is quantified in Fig. \ref{fig:sensitivity}.}
\label{tab:summary}
\begin{ruledtabular}
\begin{tabular}{lccccc}
Galaxy & $m_{21}$ & $+1\sigma$ & $-1\sigma$ & $N_{\rm valid}$ & $N_{\rm valid}/N_{\rm total}$ \\
\colrule
Carina      & 46.27 & 26.79 & 28.45 & 50000 & 1.000 \\
CVn~I       & 19.15 & 14.15 & 9.34 & 49962 & 0.999 \\
Draco       & 8.02 & 5.27 & 4.21 & 47743 & 0.955 \\
Fornax      & 38.93 & 27.19 & 23.75 & 49923 & 0.998 \\
Leo~I       & 42.76 & 34.49 & 24.60 & 50000 & 1.000 \\
Leo~II      & 35.90 & 30.42 & 18.57 & 50000 & 1.000 \\
Sculptor    & 33.72 & 25.73 & 21.39 & 49766 & 0.995 \\
Sextans     & 45.85 & 32.55 & 26.40 & 50000 & 1.000 \\
Ursa~Minor  & 10.39 & 8.21 & 6.27 & 48879 & 0.978 \\
\end{tabular}
\end{ruledtabular}
\end{table}

\subsection{Inverse-variance-weighted effective equilibrium mass}
\label{sec:combined}

Table \ref{tab:summary} reports a separate $m_{21}$ posterior for each of the nine dSphs, each summarized by its median and an asymmetric $68\%$ interval. It is useful, in addition to these nine individual constraints, to condense them into a single inverse-variance-weighted effective equilibrium mass and uncertainty, both to compare directly against previous work that quotes a single ``FDM mass'' scale (Sec. \ref{sec:discussion}) and as a natural summary statistic to be revisited once the residual-weighting and multi-galaxy extensions discussed in Secs. \ref{sec:robustness} and \ref{sec:discussion} are implemented.

For each galaxy $g=1,\dots,9$ we approximate its measurement as an independent Gaussian random variable with mean equal to the posterior median $m_g$ reported in Table \ref{tab:summary} and standard deviation
\begin{equation}
    \sigma_g \equiv \frac{\sigma_{g,+}+\sigma_{g,-}}{2},
    \label{eq:sigmaavg}
\end{equation}
i.e.\ the average of the $+1\sigma$ and $-1\sigma$ half-widths of its interval. We stress that this is an approximation: as Figure \ref{fig:m22hist} shows, several of the per-galaxy $m_{21}$ posteriors are visibly asymmetric and, in some cases (e.g.\ Sextans, Leo~I), broad and skewed toward high $m_{21}$, so a single symmetric Gaussian width does not capture the full shape of each posterior. Equation \eqref{eq:sigmaavg} is nonetheless the standard way of reducing a published asymmetric interval to a single effective Gaussian error for the purpose of combining measurements, and it is adequate for the illustrative weighted estimate constructed here.

Treating the nine galaxies as independent Gaussian measurements of a single, common effective equilibrium mass $m_{21}$, the joint log-likelihood is
\begin{equation}
\begin{split}
    \ln\mathcal{L}(m_{21}) &= -\frac{1}{2}\sum_{g=1}^{9}\frac{(m_g-m_{21})^2}{\sigma_g^2} + \mathrm{const.} \\
    &\equiv -\frac{1}{2}\chi^2(m_{21}) + \mathrm{const.},
\end{split}
    \label{eq:chi2}
\end{equation}
i.e.\ the standard $\chi^2$ statistic of weighted least-squares, Gaussian-error estimation  \cite[Chap.~15]{Press1992}. Setting $\partial\ln\mathcal{L}/\partial m_{21}=0$ gives the maximum-likelihood combined estimate
\begin{equation}
    \hat{m}_{21} = \frac{\displaystyle\sum_{g} m_g\,\sigma_g^{-2}}{\displaystyle\sum_{g}\sigma_g^{-2}},
    \label{eq:weightedmean}
\end{equation}
the familiar inverse-variance weighted mean of the nine per-galaxy medians, while the curvature of $\ln\mathcal{L}$ about its maximum, $-\partial^2\ln\mathcal{L}/\partial m_{21}^2 = \sum_g \sigma_g^{-2}$, gives the corresponding combined uncertainty
\begin{equation}
    \sigma_{\hat{m}_{21}} = \left(\sum_{g}\sigma_g^{-2}\right)^{-1/2}.
    \label{eq:weightedsigma}
\end{equation}
Equations \eqref{eq:weightedmean}-- \eqref{eq:weightedsigma} are the standard weighted-least-squares combination formulae for independent Gaussian measurements with known uncertainties, as given, e.g., in the treatment of modeling data in Press \textit{et al.} \cite[Chap.~15]{Press1992}.

Applying Eqs. \eqref{eq:weightedmean}-- \eqref{eq:weightedsigma} to the nine entries of Table \ref{tab:summary}, using Eq. \eqref{eq:sigmaavg} for each $\sigma_g$, gives the inverse-variance-weighted effective equilibrium mass (statistical uncertainty only)
\begin{equation}
    \hat m_{21} = 13.05 \pm 3.55 \qquad \left(\hat m = (1.31\pm0.35)\times10^{-20}\,\mathrm{eV}\right).
    \label{eq:pooled}
\end{equation}
Because each galaxy is weighted by $\sigma_g^{-2}$, this effective equilibrium mass is dominated by the two most tightly-constrained galaxies in the sample, Draco ($\sigma_g=4.74$) and Ursa~Minor ($\sigma_g=7.24$), which together carry $\approx80\%$ of the total weight, despite the fact that several of the more loosely-constrained galaxies (Carina, Fornax, Leo~I, Leo~II, Sculptor, Sextans; all with $\sigma_g\gtrsim23.5$) have higher median $m_{21}$ (Table \ref{tab:summary}). The weighting is not neutral: for every galaxy $\sigma_g/m_g$ lies between $0.59$ and $0.70$, so $\sigma_g^{-2}\propto m_g^{-2}$ approximately and Eq. \eqref{eq:weightedmean} favors the lowest-mass galaxies by construction; the weight carried by Draco and Ursa~Minor reflects their small median masses as much as any superior precision. The weighted mass of Eq. \eqref{eq:pooled} should therefore be read as a variance-weighted summary of the sample -- effectively anchored by Draco and Ursa~Minor -- rather than as a simple, unweighted average of the nine per-galaxy medians, and it should be interpreted together with the caveats of Secs. \ref{sec:robustness} and \ref{sec:discussion}.

Table \ref{tab:pooled} lists, for each galaxy, the quantities entering Eqs. \eqref{eq:weightedmean}-- \eqref{eq:weightedsigma}, computed from Table \ref{tab:summary}: the symmetrized width $\sigma_g$, the fractional weight $w_g=\sigma_g^{-2}/\sum_{g'}\sigma_{g'}^{-2}$, the pull $(m_g-\hat m_{21})/\sigma_g$ relative to the weighted value, and the lower $1\sigma$ edge $m_g-\sigma_{g,-}$ to be compared with the bound $m_{21}=2.2$ of Ref. \cite{Zimmermann2025}. Draco, Ursa~Minor, and CVn~I carry $56.0\%$, $24.0\%$, and $9.1\%$ of the weight, together $89\%$; the remaining six carry between $1.4\%$ and $2.3\%$ each. All pulls satisfy $|{\rm pull}|\le1.2$ and, for the fiducial window, every lower $1\sigma$ edge lies above $m_{21}=2.2$, the smallest being Draco at $3.81$, a factor $1.7$ above the bound (our $m_{21}$ values are effective equilibrium masses, not bounds).

\begin{table}[t]
\caption{Inverse-variance-weighting statistics derived from Table \ref{tab:summary} (fiducial window). $\sigma_g=(\sigma_{g,+}+\sigma_{g,-})/2$; $w_g$ is the inverse-variance weight; pull $=(m_g-\hat m_{21})/\sigma_g$ with $\hat m_{21}=13.05$; the last column is $m_g-\sigma_{g,-}$.}
\label{tab:pooled}
\begin{ruledtabular}
\begin{tabular}{lcccc}
Galaxy & $\sigma_g$ & $w_g$ (\%) & pull & $m_g-\sigma_{g,-}$ \\
\colrule
Carina     & 27.62 & 1.6 & $+1.20$ & 17.82 \\
CVn~I      & 11.75 & 9.1 & $+0.52$ & 9.81 \\
Draco      & 4.74 & 56.0 & $-1.06$ & 3.81 \\
Fornax     & 25.47 & 1.9 & $+1.02$ & 15.18 \\
Leo~I      & 29.54 & 1.4 & $+1.01$ & 18.16 \\
Leo~II     & 24.49 & 2.1 & $+0.93$ & 17.33 \\
Sculptor   & 23.56 & 2.3 & $+0.88$ & 12.33 \\
Sextans    & 29.48 & 1.4 & $+1.11$ & 19.45 \\
Ursa Minor & 7.24 & 24.0 & $-0.37$ & 4.12 \\
\end{tabular}
\end{ruledtabular}
\end{table}

Under the same independent-Gaussian assumption as Eqs. \eqref{eq:weightedmean}-- \eqref{eq:weightedsigma}, the nine medians scatter about $\hat m_{21}$ with $\chi^2=\sum_g(m_g-\hat m_{21})^2/\sigma_g^2=7.9$ for $8$ degrees of freedom ($p=0.44$), so the data do not reject a common value. This reflects the breadth of the intervals, which limits the power of the test, rather than evidence that the galaxies share a single mass; as stated above, the per-galaxy values are equilibrium masses, not samples of one population parameter. The weighted value is controlled by the two dwarfs with the smallest uncertainties (and the lowest medians). Removing Draco raises it to $19.46\pm5.35$, removing Ursa~Minor gives $13.89\pm4.07$, and removing both gives $30.37\pm7.94$; the three most heavily weighted galaxies (Draco, Ursa~Minor and CVn~I) alone give $9.80\pm3.76$. The unweighted mean of the nine medians is $31.22$ and their median is $35.90$, both well above $\hat m_{21}$ because the weighting favors the lightest systems.
Finally, the posteriors are right-skewed: the ratio $\sigma_{g,+}/\sigma_{g,-}$ ranges from $0.94$ (Carina) to $1.64$ (Leo~II), exceeding $1.2$ for seven of the nine galaxies. The symmetrized $\sigma_g$ therefore slightly overstates the lower-side uncertainty and understates the upper-side one, which is why we quote the asymmetric intervals in Table \ref{tab:summary} and treat Eqs. \eqref{eq:weightedmean}-- \eqref{eq:weightedsigma} only as a summary statistic. Across all nine galaxies, $446\,273$ of $450\,000$ posterior draws ($99.2\%$) converged, the lowest per-galaxy fraction being $95.5\%$ (Draco).

\section{Robustness and systematics}
\label{sec:robustness}

\subsection{Force-balance residual}

\begin{figure*}[tp]
    \centering
    \includegraphics[width=0.92\textwidth]{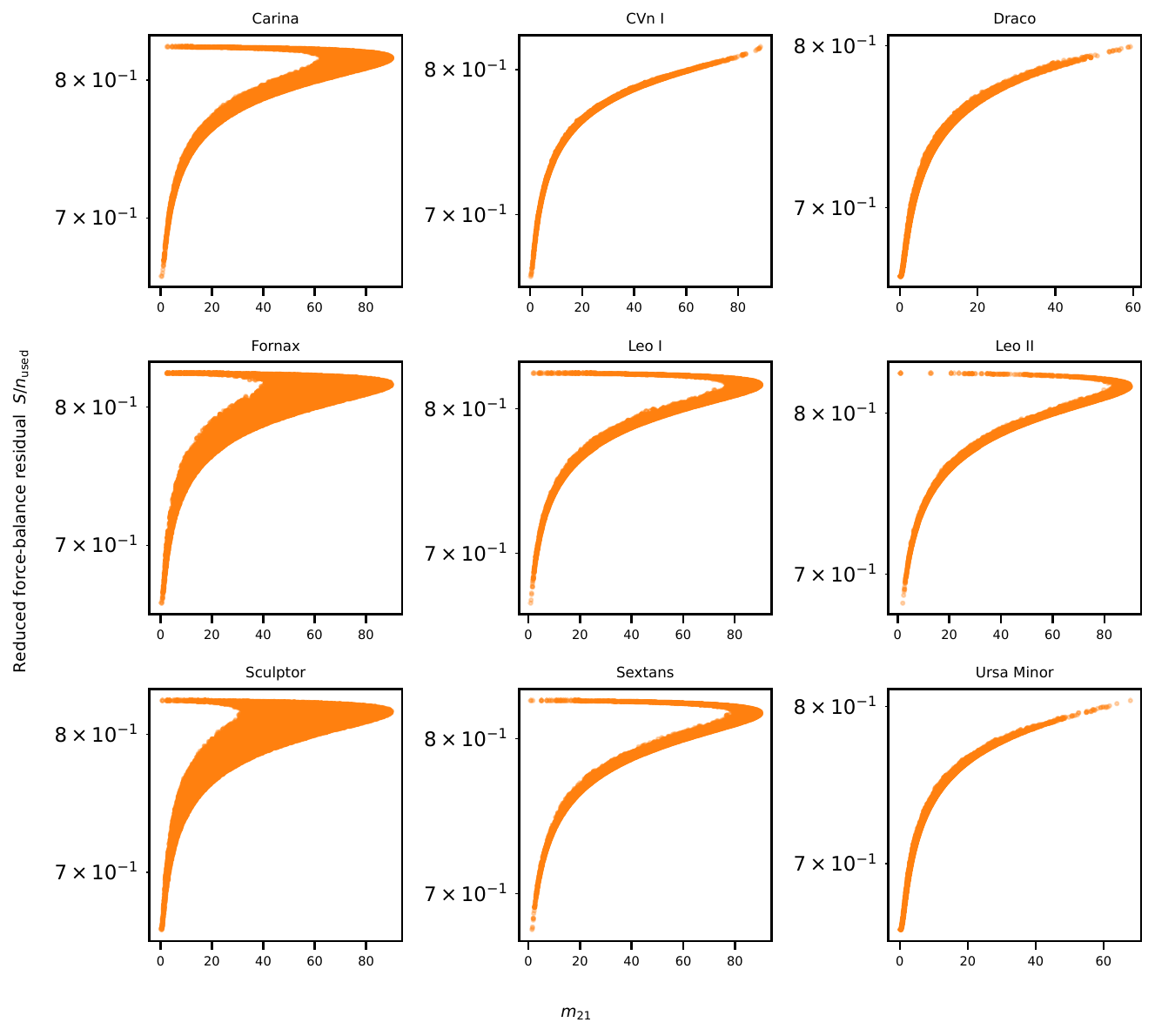}
    \caption{Fit-quality diagnostic: reduced force-balance residual $S_j/n_{{\rm used},j}$ (Eq. \ref{eq:residualsum}) plotted against the implied boson mass $m_{21}^{(j)}$, for every converged posterior draw $j$, for all nine dSphs. Each point is one \textsc{pyGravSphere} posterior realization; the horizontal axis is $m_{21}^{(j)}$ and the vertical axis is the reduced residual $S_j/n_{{\rm used},j}$ (log scale), i.e.\ the mean squared fractional force-balance mismatch of Eq. \eqref{eq:regression} averaged over the $n_{{\rm used},j}$ radial points that entered the fit for that draw; smaller values indicate a better local-equilibrium fit. Points are rasterized to keep file size manageable given the large number of posterior draws per galaxy. In every galaxy the draws with the smallest residual (the best-fitting realizations) cluster toward the low-$m_{21}$ end of the posterior, while the high-$m_{21}$ tail is associated with larger residuals. All draws have $S_j/n_{{\rm used},j}\approx0.65$--$0.83$, i.e.\ a root-mean-square fractional mismatch of order $80$--$90\%$. Because $S_j=n_{{\rm used},j}-(\sum_ix_i)^2/\sum_ix_i^2$ is invariant under a rescaling of $x_i=Q_{0,j}'(r_i)/g_j(r_i)$, it measures how much $x_i$ varies across the window, not the size of $\alpha_j$. This figure is the basis for the recommendation, in this section, that any headline $m_{21}$ constraint from this method be interpreted jointly with its residual distribution rather than in isolation.}
    \label{fig:residual}
\end{figure*}

Figure \ref{fig:residual} shows the reduced force-balance residual $S_j/n_{{\rm used},j}$ (Eq. \ref{eq:residualsum}) as a function of the implied $m_{21}^{(j)}$, for each converged posterior draw and galaxy. Two points deserve emphasis. First, the residuals are large: $S_j/n_{{\rm used},j}\approx0.65$--$0.83$ for every galaxy, i.e.\ a root-mean-square fractional mismatch of order $80$--$90\%$ between $g_j(r_i)$ and $-\alpha_jQ_{0,j}'(r_i)$ over the window. A single radius-independent mass therefore does not reproduce the force balance at the level of tens of percent for any draw, in line with the declining $m_{\rm local}(r)$ of Fig. \ref{fig:mlocal}. Second, the residual is not independent of the implied mass: draws with the smallest residual preferentially imply lower $m_{21}$, while the high-$m_{21}$ tail of each galaxy's distribution is associated with larger residuals. With the closed-form solution, $S_j=n_{{\rm used},j}-(\sum_i x_i)^2/\sum_i x_i^2$ is unchanged under a rescaling $x_i\to\lambda x_i$, so the residual depends only on how strongly $x_i=Q_{0,j}'(r_i)/g_j(r_i)$ varies across the window and not on the amplitude of $\alpha_j$. The correlation should therefore not be read as the quantum term doing little work for the heavy draws: the fit enforces $\alpha_jQ_{0,j}'\approx-g_j$ in the least-squares sense by construction. Since $m_j^2\propto|Q_{0,j}'|/g_j$, large $m_j$ arises in draws with low enclosed mass, and the correlation is consistent with such draws having a $g_j(r)$ whose radial dependence departs more strongly from the $r^{-3}$ scaling of $Q_{0,j}'$ across the window; it is a property of the reconstructed profiles. We therefore recommend that any headline $m_{21}$ value derived from this method be interpreted together with the residual distribution of Figure \ref{fig:residual}. A natural refinement, which we identify here but leave for future work (Sec. \ref{sec:discussion}), is a residual-weighted summary statistic -- for example a weighted median of $\{m_{21}^{(j)}\}$ with weights $w_j \propto \exp(-S_j/n_{{\rm used},j})$ or $w_j \propto (S_j/n_{{\rm used},j})^{-1}$ -- which would down-weight, rather than discard, the poorly-fit high-residual draws responsible for the high-$m_{21}$ tail identified above.

\subsection{Convergence fraction}

\begin{figure}[t]
    \centering
    \includegraphics[width=\columnwidth]{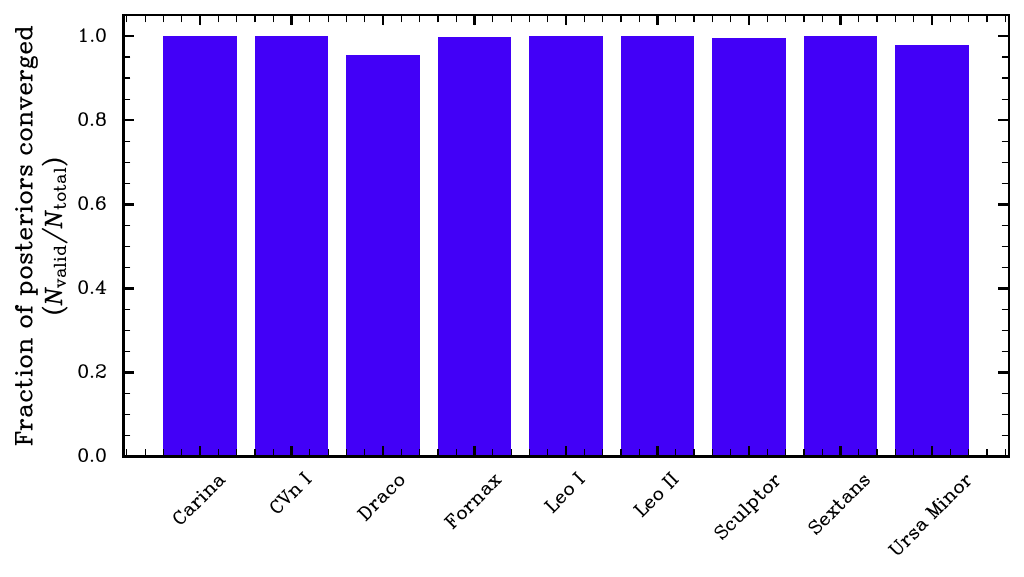}
    \caption{Convergence fraction $N_{\rm valid}/N_{\rm total}$ of the force-balance regression, one bar per galaxy, for all nine dSphs (numerical values also listed in Table \ref{tab:summary}). Here $N_{\rm total}$ is the total number of \textsc{pyGravSphere} posterior draws available for a given galaxy ($N_{\rm total}=50000$ for every galaxy in our sample) and $N_{\rm valid}$ is the number of those draws for which the per-draw regression of Eq. \eqref{eq:regression} converged, i.e.\ for which at least twenty five radial points within the fiducial window $[0.05,1.0]\,\kpc$ satisfied the required sign combination ($g_j(r_i)>0$, $Q_{0,j}'(r_i)<0$) and the resulting $\alpha_j$ was positive (Sec. \ref{sec:method}). A convergence fraction close to unity indicates that essentially the full \textsc{pyGravSphere} posterior chain contributes to that galaxy's $m_{21}$ distribution in Fig. \ref{fig:m22hist}; a lower fraction means the reported posterior is built from a smaller subset of the full \textsc{pyGravSphere} chain and should be interpreted with correspondingly greater caution, since that subset need not be statistically representative of the complete posterior. In our sample all nine galaxies converge for at least $95\%$ of their posterior draws, with Draco (the lowest, $95.5\%$) and Ursa~Minor ($97.8\%$) showing the mildest reductions relative to Carina, Leo~I, Leo~II, and Sextans, for which the regression converges on effectively every draw ($100.0\%$).}
    \label{fig:convergence}
\end{figure}

Figure \ref{fig:convergence} shows the fraction of posterior draws per galaxy that satisfied our convergence criteria ($n_{\rm used}\ge25$ radial points with the correct sign combination, and $\alpha_j>0$). A low convergence fraction for a given galaxy indicates that its reported $m_{21}$ posterior is built from a minority, and potentially non-representative, subset of the full GravSphere posterior chain, and should be weighted accordingly when comparing across galaxies. We note that the convergence criteria are sign conditions on $g_j$, $Q_{0,j}'$ and $\alpha_j$ only; a convergence fraction close to unity therefore says nothing about the quality of the force-balance fit, for which the residuals of Fig. \ref{fig:residual} must be consulted.

\subsection{Sensitivity to the radial fitting window}

\begin{figure*}[tp]
    \centering
    \includegraphics[width=0.92\textwidth]{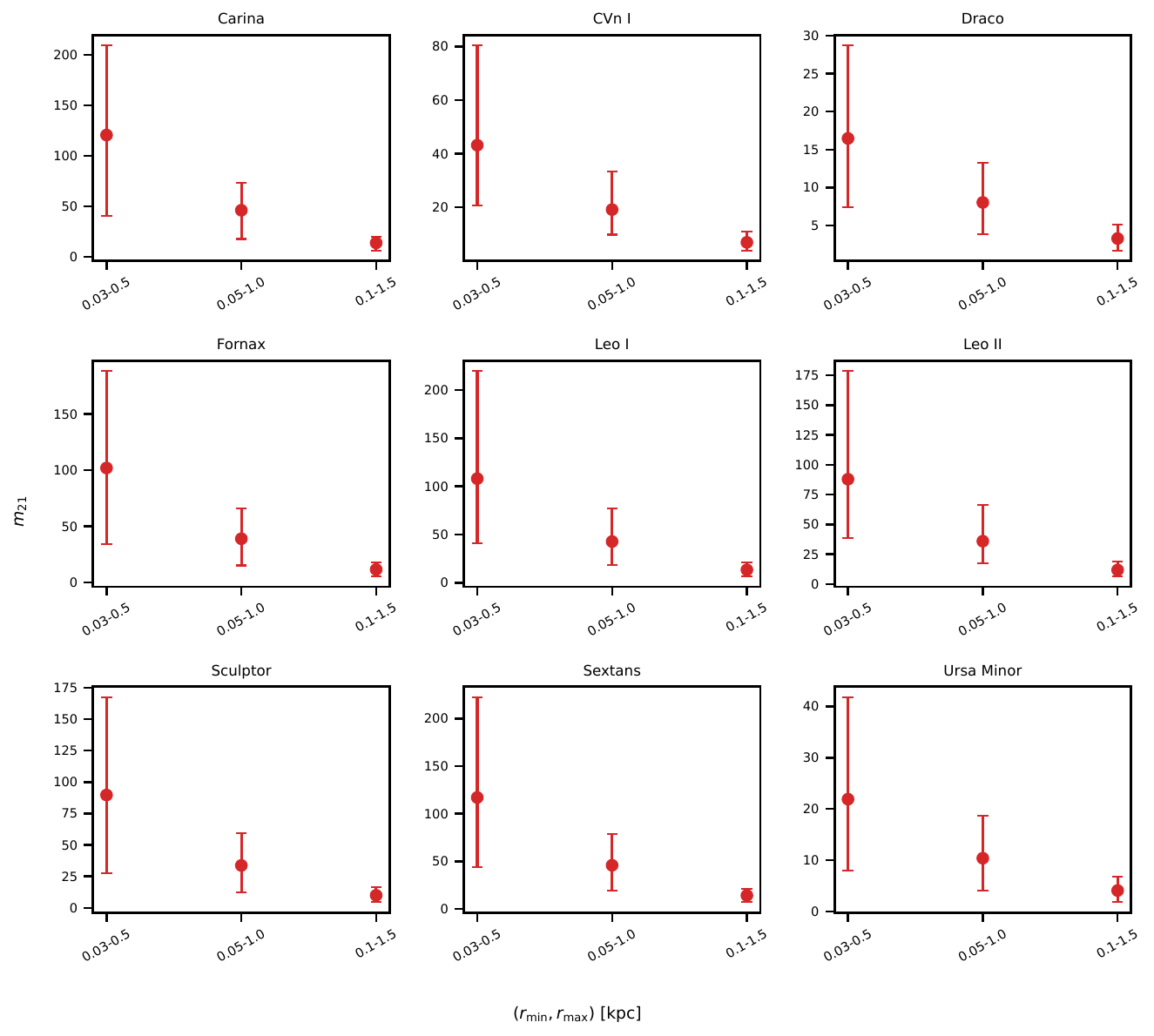}
    \caption{Sensitivity of the recovered boson mass to the choice of radial fitting window, for all nine dSphs. For each galaxy we show three points (offset horizontally for legibility): the posterior median $m_{21}$ (with vertical error bar spanning the $68\%$, $16$th--$84$th-percentile interval) recovered when the force-balance regression of Eq. \eqref{eq:regression} is applied over three different radial windows $(r_{\min},r_{\max})$ -- a narrow, inner-weighted window $(0.03,0.5)\,\kpc$; the fiducial window $(0.05,1.0)\,\kpc$ used for the headline results of Table \ref{tab:summary} and Figs. \ref{fig:m22hist}-- \ref{fig:summary}; and a wider, outer-extended window $(0.1,1.5)\,\kpc$ -- with all other convergence criteria held fixed. Galaxies for which the three points agree within their $68\%$ intervals are those for which the equilibrium-mass constraint is robust to this methodological choice; galaxies showing a significant systematic shift between windows indicate that the assumption of a single, radius-independent boson mass is only approximately valid over the full radial range probed, and for such galaxies we recommend quoting the window-to-window spread as an additional systematic uncertainty alongside the statistical interval of Table \ref{tab:summary} (this section).}
    \label{fig:sensitivity}
\end{figure*}

Because the fiducial window $(r_{\min},r_{\max})=(0.05,1.0)\,\kpc$ used in Eq. \eqref{eq:regression} is not derived from first principles, we repeat the regression for two alternative windows, $(0.03,0.5)\,\kpc$ and $(0.1,1.5)\,\kpc$, and compare the resulting median $m_{21}$ per galaxy in Figure \ref{fig:sensitivity}. As Fig. \ref{fig:sensitivity} shows, for every galaxy the median $m_{21}$ changes by up to roughly an order of magnitude between these windows, far more than the statistical intervals of Table \ref{tab:summary}, with narrower inner windows giving heavier masses. The window dependence is therefore a dominant systematic uncertainty, the values of Table \ref{tab:summary} and the weighted mass of Eq. \eqref{eq:pooled} are conditional on the fiducial window, and we recommend that this dependence be quoted as an additional systematic uncertainty whenever the method is applied.


\subsection{Summary of diagnostics and limitations}
\label{sec:diagnostics}

Three diagnostics qualify how Table \ref{tab:summary} should be read.

(i)\emph{Local estimator.} A radially flat $m_{\rm local}(r)$ would confirm a single equilibrium mass. Instead, for every galaxy $m_{\rm local}(r)$ declines smoothly and monotonically by roughly one and a half to two and a half orders of magnitude over $r\approx0.05$--$1\,\kpc$ (Fig. \ref{fig:mlocal}). This is the generic expectation for a profile that is not a soliton: for $\rho\propto r^{-\gamma}$, Eq. \eqref{eq:Qform} gives $Q_0'\propto r^{-3}$ almost independently of $\gamma$, whereas $g$ follows the enclosed mass. It reflects how far the reconstructed shape departs from a soliton of fixed mass, and possibly a window that extends beyond the coherent soliton core into the granular, non-static envelope  \cite{Schive2014a,Veltmaat2018}, or growing effects of flows, non-sphericity, and tides at large $r$. Equation \eqref{eq:regression} therefore returns an \emph{effective, window-dependent} equilibrium mass rather than a mass independently confirmed at each radius.

(ii)\emph{Residual--mass correlation.} In every galaxy the draws with the smallest residual preferentially imply low $m_{21}$, while the high-$m_{21}$ tail is populated by draws with larger residuals (Fig. \ref{fig:residual}). The residuals are large for all draws ($S_j/n_{{\rm used},j}\approx0.65$--$0.83$), and since $S_j$ is invariant under a rescaling of the regressors the correlation reflects the shape of the reconstructed profiles, not the contribution of the quantum term. Downweighting high-residual draws, e.g.\ with weights $w_j\propto(S_j/n_{{\rm used},j})^{-1}$, would lower the medians; we leave this to future work.

(iii)\emph{Fitting window.} Repeating the regression for $(r_{\min},r_{\max})=(0.03,0.5)$ and $(0.1,1.5)\,\kpc$ changes the medians by up to roughly an order of magnitude, far more than the statistical intervals of Table \ref{tab:summary}, with narrower inner windows giving heavier masses (Fig. \ref{fig:sensitivity}). Window dependence should be carried as a dominant systematic uncertainty, and the values of Table \ref{tab:summary} are conditional on the fiducial window.

\section{Discussion}
\label{sec:discussion}

Our per-galaxy equilibrium-mass measurements can be compared, galaxy by galaxy, against independent FDM boson-mass constraints derived using different methodology and, in some cases, different galaxies. Gonz\'alez-Morales \textit{et al.} \cite{GonzalezMorales2017} caution that na\"ive Jeans-based axion-mass inferences from dSphs are subject to significant mass-anisotropy-degeneracy bias unless higher-order kinematic information (as incorporated in GravSphere) is used; our use of \textsc{pyGravSphere} posteriors directly inherits this mitigation. Chen \textit{et al.} \cite{Chen2017} performed a dedicated wave-dark-matter Jeans analysis of dSphs and found soliton-core solutions broadly consistent with $m_{21} \sim \mathcal{O}(0.1)$ for several systems in our sample; our median values ($m_{21}\sim8$--$46$, Table \ref{tab:summary}) are systematically one to two orders of magnitude heavier, a discrepancy that could reflect the different physical assumption underlying the two methods (a fully relaxed, parametric soliton profile in Ref. \cite{Chen2017} versus our purely local, non-parametric equilibrium condition), the residual-mass correlation identified in Sec. \ref{sec:robustness}, or the radially declining $m_{\rm local}(r)$ identified in Sec. \ref{sec:results} -- which indicates that our fitting window may extend beyond the coherent soliton core into the non-equilibrium envelope of the halo, biasing $m_{21}$ upward relative to a genuine core-equilibrium measurement -- and merits further investigation. A direct test, which we defer to future work together with the residual-weighted statistic proposed in Sec. \ref{sec:robustness}, is to restrict the regression to the flattest sub-region of $m_{\rm local}(r)$ identified on a per-galaxy basis and check whether this narrows the gap with the soliton-fit-based results of Chen \textit{et al.} \cite{Chen2017}. Robles \textit{et al.} \cite{RoblesBullockBoylanKolchin2019} similarly caution that the soliton-halo relation implied by scalar-field dark matter simulations does not, in general, uniquely fix a single galaxy-independent boson mass once realistic halo-to-halo scatter is included, which may partly explain the galaxy-to-galaxy spread we observe. Zimmermann \textit{et al.} \cite{Zimmermann2025} obtained a model-independent lower bound $m>2.2\times10^{-21}\,\mathrm{eV}$ ($m_{21}>2.2$) from Leo~II alone using a full phase-space reconstruction, a substantially more data-intensive approach than the force-balance regression used here; our Leo~II median, $m_{21}=35.90^{+30.42}_{-18.57}$, comfortably exceeds this bound for the fiducial window even at its $1\sigma$ lower edge, so the two methods -- despite resting on very different assumptions -- agree on the qualitative conclusion that Leo~II favors a boson mass well above $10^{-21}\,\mathrm{eV}$. Dalal and Kravtsov \cite{Dalal2022} used the sizes and kinematics of ultra-faint dwarfs to exclude $m\lesssim3\times10^{-19}\,\mathrm{eV}$. Our equilibrium masses lie below this value, so if FDM makes up all of the DM and their size--mass argument applies, the two results are in tension. The analyses probe different galaxy populations and rest on different physical assumptions, and our estimator returns an effective equilibrium mass rather than a bound, but this tension should be resolved by future work. Independent cosmological and astrophysical probes place complementary constraints spanning a wide range of scales and methodologies: CMB and large-scale-structure analyses  \cite{Hlozek2015}, the Lyman-$\alpha$ forest  \cite{Armengaud2017,Irsic2017,Kobayashi2017,RogersPeiris2021}, the Milky Way satellite luminosity function  \cite{Nadler2021}, reionization-era galaxy counts  \cite{Bozek2015}, and dedicated ultrafaint-dwarf analyses  \cite{MarshNiemeyer2019,HayashiFerreiraChan2021} together disfavor $m\lesssim$ a few $\times10^{-21}\,\mathrm{eV}$, a range below all of our per-galaxy medians. The strongest of these, the Lyman-$\alpha$ forest bound of Rogers and Peiris \cite{RogersPeiris2021}, $m>2\times10^{-20}\,\mathrm{eV}$ at $95\%$ confidence, lies above our inverse-variance-weighted effective mass, $(1.31\pm0.35)\times10^{-20}\,\mathrm{eV}$, and above the upper $68\%$ limits of Draco and Ursa~Minor, the two galaxies that dominate that weighted value, while six of the nine medians in Table \ref{tab:summary} lie above it. The window dependence of Sec. \ref{sec:robustness} is much larger than the factor of about $1.5$ separating the weighted mass from this bound, so we do not regard this as a quantitative discrepancy, but it shows that the weighted mass should not be quoted as a measurement of the cosmological boson mass. We also caution that these probes constrain the cosmological boson mass under the assumption that FDM constitutes all of the dark matter, whereas our method constrains only the local equilibrium mass implied by nine individual halos.

A further consistency question concerns the regime of validity of the static, curl-free equilibrium itself. For $m_{21}\gtrsim10$ the de~Broglie wavelength $\hbar/(mv)$ at a typical dSph velocity $v\sim10\,\mathrm{km\,s^{-1}}$ is of order $20\,\mathrm{pc}$, and the soliton core radius found in cosmological simulations \cite{Schive2014a,Veltmaat2018}, which scales as $(m\,M_h^{1/3})^{-1}$, is of the same order ($\sim10$--$20\,\mathrm{pc}$) for halo masses of $\sim10^{9}\,\msun$. Both lie below the inner edge of the fiducial window, $0.05\,\kpc$. If the equilibrium masses of Table \ref{tab:summary} were the true boson mass, the window would therefore lie largely in the granular, wave-interference-supported envelope, where the static, curl-free balance of Eq. \eqref{eq:balance} is not expected to hold at any instant. This is a further reason to regard $m_{21}$ as an effective equilibrium mass of the reconstructed profile and not as a measurement of the fundamental boson mass.

We emphasize, as noted in Sec. \ref{sec:theory} and Sec. \ref{sec:robustness}, that our method measures the boson mass \emph{required for the observed density and mass profile of each galaxy to be in local static equilibrium}, rather than a population-level statistical bound on a single true value of $m$ shared by all dSphs. The weighted value of Eq. \eqref{eq:pooled} assumes a single $m$ shared by all nine halos and is not a population-level bound. A galaxy with a poor equilibrium fit (high residual, Fig. \ref{fig:residual}) is evidence either that the static, curl-free, spherical configuration is not realized in that system, or that its $M(<r)$ and $\rho(r)$ are in tension with FDM equilibrium at any boson mass. A natural extension of this work, which we defer to future study, is to test for a correlation between the per-galaxy equilibrium mass and independent galaxy properties (e.g., central density, core radius, or orbital pericenter, which sets the strength of Milky-Way tidal perturbation), which could help distinguish genuine astrophysical variation from a common origin.

In future work, we plan to investigate the role of self-interactions and test whether this additional parameter is required in the FDM model. Our preliminary study shows that including an attractive self-interaction improves the fit per degree of freedom. We find that a positive (attractive) interaction is favoured across all the dSphs \cite{MitraDutta_m_lambda}.

\section{Conclusions}
\label{sec:conclusion}

We have introduced and applied a direct, per-posterior-draw regression method for constraining the fuzzy dark matter boson mass from the requirement of local hydrostatic equilibrium between classical gravity and the SP--Madelung quantum-pressure gradient, using \textsc{pyGravSphere} density and mass posteriors for nine Milky Way dSphs. Unlike parametric soliton-fitting approaches, our estimator makes no assumption about the functional form of $\rho(r)$ beyond the piecewise power-law binning of the \textsc{pyGravSphere} reconstruction, spherical symmetry and a locally static velocity field, and it yields a distribution of implied boson masses for each galaxy individually, propagating the kinematic uncertainty of the reconstruction. We find per-galaxy median boson masses in the range $m_{21}\sim8$--$46$ across the nine dSphs (Table \ref{tab:summary}), with Draco ($m_{21}=8.02^{+5.27}_{-4.21}$) and Ursa~Minor ($m_{21}=10.39^{+8.21}_{-6.27}$) favoring the lightest boson masses in the sample and Carina ($m_{21}=46.27^{+26.79}_{-28.45}$) and Sextans ($m_{21}=45.85^{+32.55}_{-26.40}$) the heaviest, with the remaining five galaxies (CVn~I, Fornax, Leo~I, Leo~II, Sculptor) spanning intermediate values $m_{21}\sim19$--$43$. For the fiducial window, all nine median values, including their $1\sigma$ lower edges, lie above the independent lower bound $m>2.2\times10^{-21}\,\mathrm{eV}$ ($m_{21}>2.2$) of Zimmermann \textit{et al.} \cite{Zimmermann2025}; since our quantities are effective equilibrium masses and not bounds, this is a statement of qualitative consistency. Combining the nine per-galaxy results into an inverse-variance-weighted effective equilibrium mass (Sec.  \ref{sec:combined}) gives $\hat m_{21}=13.05\pm3.55$ (statistical uncertainty only), i.e.\ $\hat m=(1.31\pm0.35)\times10^{-20}\,\mathrm{eV}$, dominated by Draco and Ursa~Minor, whose small medians (the uncertainties scale with the medians) give them the largest weights. We find that the residuals of the force-balance fit are large (a root-mean-square fractional mismatch of order $80$--$90\%$) and correlated with the implied mass, and that the results depend strongly on the radial fitting window: the per-galaxy medians change by up to roughly an order of magnitude between the windows tested (Fig. \ref{fig:sensitivity}), far more than the statistical intervals. This window dependence is the dominant systematic uncertainty and should be included in the error budget whenever this method is applied to additional systems. These per-galaxy effective equilibrium masses are a diagnostic of the reconstructed profiles that is complementary to existing statistical and phase-space-based dSph constraints on the FDM boson mass; they are not themselves bounds on it.
\section*{Data and Code Availability}
The analysis pipeline used in this work -- the force-balance regression notebook, the figure-generation script, and the summary tables underlying Table \ref{tab:summary} and Figs. \ref{fig:rho}-- \ref{fig:sensitivity} -- will be made available at a public repository upon publication (or earlier upon reasonable request to the corresponding author). The underlying \textsc{pyGravSphere} density and enclosed-mass posterior chains for the nine dSphs are available from the \textsc{pyGravSphere} repository  \cite{pyGravSphere}, and the resolved stellar-kinematic datasets from which they are derived are publicly available from the original sources cited in Sec. \ref{sec:data}; the CVn~I kinematic data are additionally available, with usage instructions, from the data repository maintained by J.\ Simon at \url{https://users.obs.carnegiescience.edu/jsimon/data.html}.


\begin{acknowledgments}
We would like to thank Gautam Das (IISER Kolkata) for emcee discussions, Bihag Dave (IIT Madras) for useful, insightful discussions on FDM and Prof. Ritesh Kumar Singh (IISER Kolkata) on useful feedback on statistical analysis. We gratefully acknowledge the use of the \textsc{pyGravSphere} package  \cite{pyGravSphere}, which builds on the core \texttt{GravSphere} code  \cite{ReadSteger2017,ReadDraco2018,Genina2020,Collins2021}, available at \url{https://github.com/justinread/gravsphere}, and express our sincere thanks to Dr. Anna Genina for her work developing \textsc{pyGravSphere} as an independent, ready-made public wrapper for the core \texttt{GravSphere} code. \textsc{pyGravSphere} relies on the \texttt{emcee} affine-invariant ensemble sampler  \cite{ForemanMackey2013} for its Markov-chain posterior estimation. S.M. acknowledges financial support from the University Grants Commission (UGC), India.
Finally, the authors acknowledge the support and infrastructure provided by the Department of Physical Sciences at the Indian Institute of Science Education and Research (IISER) Kolkata. We used Claude Sonnet 5.5 and ChatGPT 6 for English-language editing and
to improve the readability and fluency of the manuscript.
\end{acknowledgments}

\bibliography{referencesv2}

\end{document}